\documentclass[amsmath,amssymb,11pt]{article}
\usepackage{jheppub2}
\pdfoutput=1
\usepackage{graphicx,epsfig}
\usepackage{float}
\usepackage{subfig}
\usepackage{subfloat}
\usepackage[utf8]{inputenc}
\usepackage{hyperref}
\usepackage{caption}
\usepackage{color}

\usepackage{tikz}
\usetikzlibrary{arrows}

\def\t{\tau}

\newcommand*{\affmark}[1][*]{\textsuperscript{#1}}

\newcommand{\beq}{\begin{equation}}
\newcommand{\eeq}{\end{equation}}
\newcommand{\be}{\begin{equation}}
\newcommand{\ee}{\end{equation}}

\subheader{\begin{flushright}
\texttt{IFT-UAM/CSIC-22-64\\
UTTG-09-2022}
\end{flushright}}

\title{Membrane nucleation rates from holography}

\author{Maite Arcos,\affmark[a]}
\emailAdd{maite.arcos@ucl.ac.uk}
\author{Willy Fischler,\affmark[b]}
\emailAdd{fischler@physics.utexas.edu}
\author{Juan F. Pedraza,\affmark[c]}
\emailAdd{j.pedraza@csic.es}
\author{and Andrew Svesko\affmark[a]}
\emailAdd{a.svesko@ucl.ac.uk}

\affiliation{\affmark[a]Department of Physics and Astronomy, University College London, London, WC1E 6BT, UK\\
\affmark[b]Department of Physics, The University of Texas at Austin,
Austin, Texas 78712, USA\\
\affmark[c]Instituto de F\'isica Te\'orica UAM/CSIC, Calle Nicol\'as Cabrera 13-15, Madrid 28049, Spain}

\abstract{Membrane nucleation, a higher dimensional analog of the Schwinger effect, is a useful toy model for vacuum decay. While a non-perturbative effect, the computation of nucleation rates has only been accomplished at weak coupling in the field theory. Here we compute the nucleation rates of spherical membranes using AdS/CFT duality, thus naturally including the effects of strong coupling. More precisely, we consider the nucleation of spherical membranes coupled to an antisymmetric tensor field, a process
which renders the vacuum unstable above a critical value of the field strength.  We analyze membrane creation in flat and de Sitter space using various foliations of AdS. This is accomplished via instanton methods, where the rate of nucleation is dominated by the semi-classical on-shell Euclidean action. Our findings generalize the holographic Schwinger effect and provide a step toward holographic false vacuum decay mediated by Coleman-De Luccia instantons.}

\begin{document}

\maketitle

\section{Introduction} \label{sec:intro}

Quantum tunneling, a non-perturbative phenomenon for which there is no classical counterpart, plays an essential role in a number of physical processes.  A famous example of tunneling is the decay of the false vacuum due to the nucleation of bubbles of true vacuum during first order phase transitions \cite{Coleman:1977py,Callan:1977pt}. Consequently, a curved de Sitter spacetime -- an idealized approximation of our universe in the far past and future -- is possibly in a metastable false vacuum state \cite{Coleman:1980aw}.\footnote{Alternatively, via the creation of false vacuum bubbles, a de Sitter-like background may undergo true vacuum decay \cite{Lee:1987qc}.}  Additional fundamental processes with a tunneling description range from nuclear fusion to Hawking radiation \cite{Parikh:1999mf}, to cosmological phase transitions and structure formation, e.g., \cite{Hiscock:1987hn,Gregory:2013hja,Burda:2015isa}.  A particularly interesting subclass of  tunneling processes  include the nucleation of membranes in flat and curved backgrounds.  As a higher dimensional analog of Schwinger pair production \cite{Schwinger:1951nm}, membrane nucleation provides a mechanism for the neutralization of the cosmological constant \cite{Brown:1987dd,Brown:1988kg} (see \cite{Kaloper:2022oqv,Kaloper:2022utc,Kaloper:2022jpv} for a modern treatment), and acts as an illustrative precursor for studying the materialization of various topological defects during cosmic inflation \cite{Garriga:1993fh}.

By now it is standard practice to describe tunneling processes using instanton techniques. Instantons, as classical solutions to the Euclidean equations of motion obeying suitable boundary conditions, are saddle points to a Euclidean path integral and therefore are the leading order contribution to the path integral in a saddle-point approximation. The leading order contribution to the nucleation rate $\Gamma$, or, equivalently, the exponential behavior of the tunneling probability, of an object is found by computing the partition function and is thereby proportional to the exponential of the on-shell Euclidean action $S_{E}$ of the system of interest (see \cite{Coleman:1985rnk} for a pedagogical review)
\beq \Gamma=Ae^{-S_{E}}\;.\label{eq:nucrategenintro}\eeq
Here the prefactor $A$ is typically comprised of functional determinants generated by integrating out small quantum fluctuations about the instanton stationary point, thus encoding typically subleading 1-loop quantum corrections.
For example, the Schwinger nucleation rate of a spin-$j$ particle-antiparticle pair of mass $m$ in a strong electric field $E$ is \cite{Schwinger:1951nm}
\beq \Gamma=\frac{(2j+1)E^{2}}{8\pi^{3}}\sum_{n=1}^{\infty}\frac{(-1)^{(n+1)(2j+1)}}{n^{2}}e^{-\frac{\pi m^{2}n}{E}}\;,\label{eq:Schwingerrate}\eeq
which can be recovered from a sum over instanton amplitudes to tunnel through a potential barrier \cite{Affleck:1981bma}. Similar derivations hold for the computation of nucleation rates of strings, membranes, and topological defects, in both flat and curved backgrounds \cite{Basu:1991ig,Garriga:1993fh,Garriga:1994bm}.

Despite being non-perturbative in nature, an unfortunate limitation of each of the aforementioned computations, including the rate (\ref{eq:Schwingerrate}), directly or indirectly assume the quantum field theory is weakly coupled. Of course, weakly coupled field theories comprise only a small class among the landscape of field theories; indeed the standard model of particle physics is an example of a strongly coupled field theory. It is of interest then to compute nucleation rates beyond the weak coupling regime. In this article we take steps in addressing this question and analyze nucleation rates of spherical membranes in flat and de Sitter backgrounds using the Anti-de Sitter space/conformal field theory (AdS/CFT) correspondence.

AdS/CFT, born from studies in string theory, is a non-perturbative candidate model of quantum gravity, in which gravitational physics in a (bulk) $d+2$-dimensional AdS background has a dual description in terms of a conformal field theory living on the $d+1$-dimensional conformal boundary of AdS. The correspondence is thus a specific realization of the holographic principle. A striking feature of the AdS/CFT correspondence is that it is a strong-weak coupling duality: coupling constants between the bulk and boundary theories are inversely related, such that strongly coupled field computations on the boundary may instead be performed via a gravity calculation at weak coupling. For example, the challenging task of computing entanglement entropy of certain strongly coupled field theories can be accomplished by computing the area of bulk minimal surfaces obeying certain homology conditions \cite{Ryu:2006bv}. Further, AdS/CFT was used to predict a universal value of the shear viscosity to entropy density ratio in a wide class of theories with classical gravity duals \cite{Kovtun:2004de}, in remarkable agreement with the value measured experimentally for the quark-gluon plasma produced in heavy-ion collision experiments \cite{Luzum:2008cw}. Thus, at the very least, AdS/CFT has proven to be a powerful tool to analyze aspects of strongly coupled field theories. Here we exploit this feature  to compute the nucleation rates of membranes at strong coupling.

Progress along these lines has already been made. In particular, the Schwinger pair production rate of W-bosons in a supersymmetric Yang-Mills theory in Minkowski space was computed using holographic methods, leading to a holographic generalization of Schwinger's pair production formula (\ref{eq:Schwingerrate}) including strong coupling effects \cite{Semenoff:2011ng}. From the bulk perspective, this amounts to evaluating the on-shell Euclidean action of a string ending on a stack of flavor branes. The endpoints of the string are coupled to a Maxwell field with a constant electric field, and represent the particle-antiparticle Schwinger pair.  The holographic Schwinger effect has since been extended in a variety of ways, e.g., \cite{Ambjorn:2011wz,Sato:2013pxa,Sato:2013iua,Kawai:2013xya,Wu:2015krf},  including the study of pair production in curved backgrounds, non-relativistic backgrounds and confining models, among others \cite{Sato:2013hyw,Fischler:2014ama,BitaghsirFadafan:2015asm,Kawai:2015mha,Ghodrati:2015rta,Shahkarami:2015qff}. In particular, the rate of production of Schwinger pairs in de Sitter space was achieved by considering de Sitter foliations of $\text{AdS}$ \cite{Fischler:2014ama}, extending various non-holographic computations to the strong coupling regime \cite{Garriga:1993fh,Garriga:1994bm,Villalba:1995za,Kim:2008xv,Stahl:2015gaa,Bavarsad:2016cxh,Stahl:2016geq,Sharma:2017ivh,Grewal:2021bsu}.

In this article we move beyond the holographic Schwinger effect and use AdS/CFT to study the nucleation of spherical membranes in  flat and de Sitter backgrounds of arbitrary spacetime dimension. More precisely, we consider the creation of spherical membranes coupled to a higher rank antisymmetric tensor field, which can be understood as a toy model for false vacuum decay. The nucleation rate and critical value of the field strength are computed by evaluating the on-shell membrane action. Our results thus extend the aforementioned holographic Schwinger effect, reducing to this case in the appropriate limit, as well as a strong coupling generalization of conventional field theoretic constructions. Of note is the evaluation of the on-shell Euclidean  membrane action in de Sitter space which is determined completely analytically in arbitrary dimensions.

The remainder of this article is as follows. In Section \ref{sec:setupreview} we briefly review the holographic set-up and computation of the on-shell action of a string used in the holographic Schwinger effect. Section \ref{sec:flatrates} is devoted to the analysis of membrane nucleation rates in flat space, by computing the on-shell action for a membrane ending on a cutoff surface near the flat conformal boundary of $\text{AdS}_{d+2}$. We find the Lorentzian continuation of the instanton describes a spherical membrane contracting and expanding at constant proper acceleration, and relate the rest energy of the membrane to the bulk cutoff. Similarly, in Section \ref{sec:dSrates} we compute the membrane nucleation rate in de Sitter space by considering various $(d+1)$-dimensional de Sitter space foliations of AdS. In this case and in the limit the field strength vanishes, our results yield the spontaneous nucleation rate for defects solely due to the background inflation. We summarize and provide concluding remarks in Section \ref{sec:conclusion}, discussing multiple future avenues worth exploring. For the sake of completeness we include two appendices. Appendix \ref{app:nucweak} reviews the computation of membrane nucleation rates using conventional quantum field theory at weak coupling \cite{Garriga:1993fh}, and Appendix \ref{app:dSfoliationsAdS} lists multiple de Sitter foliations of AdS.

%%%%%%%%%%%%%%%%%%%%%%%%%%%%%%%%%%%%%%%%%%%%%%%%%%%%%%%%%%

\section{Set-up and review of holographic Schwinger effect} \label{sec:setupreview}

We begin by providing an elementary review of the holographic set-up used to study the holographic Schwinger effect, which we analyze using instanton methods by computing the on-shell Nambu-Goto action for a string coupled to a Maxwell field. This review will act as a useful comparison when we consider the nucleation of spherical membranes as many of the techniques generalize in a straightforward manner.

\subsection{Gravity set-up: preliminaries}

Broadly, the AdS/CFT correspondence says the following: the states of certain conformal field theories in the large $N_c$ limit and large 't Hooft coupling $\lambda=g_{\text{YM}}^2N_c$ living on a $d+1$ dimensional spacetime $\mathcal{M}_{d+1}$ are dual to solutions of supergravity theories which asymptotically approach $\text{AdS}_{d+2}\times X$, where $X$ is a compact manifold whose isometries are recognized as the global symmetries of the field theory, and the boundary of $\text{AdS}_{d+2}$ is identified with $\mathcal{M}_{d+1}$. In its strongest form, the duality purports a full equivalence between the partition functions between the boundary CFT and the bulk theory. However, for practical purposes it is often discussed within a saddle-point approximation, valid at large $N_c$, because bulk calculations are tractable in this regime \cite{Gubser:1998bc,Witten:1998qj}. The canonical example of AdS/CFT duality relates type IIB string theory on an $\text{AdS}_{5}\times S^{5}$ background with $N_{c}$ units of Ramond-Ramond flux through $S^{5}$ to $\mathcal{N}=4$ $SU(N_{c})$ super-Yang-Mills theory, though we will not restrict ourselves to a specific version of the correspondence.

In fact, it is often sufficient to consider a bulk spacetime $\mathcal{B}_{d+2}\times X$ for an appropriate internal space $X$, where $\mathcal{B}_{d+2}$ is a solution to Einstein's equations with a negative cosmological constant $\Lambda=-\frac{d(d+1)}{2L^{2}}$, where $L$ being the length scale of the asymptotic $\text{AdS}_{d+2}$ spacetime.
In this context one relates the bulk $(d+2)$-dimensional Newton's constant $G_{N}$ to the central charge $c$ of the dual CFT via
\beq c=\frac{L^{d}}{16\pi G_{N}}\;.\label{eq:centchargec}\eeq
Hence, for fixed radius $L$, studying holographic CFTs in the large $c$ limit amounts to considering (classical) gravity in the $G_{N}\to0$ limit, neglecting quantum gravity corrections. With this in mind, we will consider the case when the bulk spacetime is represented by empty $\text{AdS}_{d+2}$, which in Poinca\'re coordinates has the line element
\be
ds^2_{d+2}=\frac{L^2}{z^2}\left(-dt^2+d\vec{x}^2+dz^2\right)\,,
\label{eq:flatads2back}\ee
where $d\vec{x}^{2}=dx_{1}^{2}+dx_{2}^{2}+...dx_{d}^{2}$. According to the standard AdS/CFT dictionary, this geometry is dual to the vacuum of a $(d+1)$-dimensional CFT with central charge $c$ (\ref{eq:centchargec}). Here the coordinate $z$ is a bulk `radial' coordinate, such that the (flat) conformal boundary is located at $z=0$. In the following two sections we will consider string and membrane solutions living in this background subject to specific boundary conditions.

As described above, Schwinger pair production or membrane nucleation requires one to couple to a Maxwell gauge field or a higher antisymmetric tensor field, respectively. This means one needs to add fundamental matter to the field theory. Holographically this would entail introducing a stack of $N_{f}$ flavor branes in the bulk geometry. With the canonical example in mind, when $N_{f}\ll N_{c}$, the flavor branes act as probe branes, i.e., we may neglect their backreaction effects on the background geometry and treat the geometry as given. Furthermore, the flavor branes will be assumed to expand in all directions on the boundary, however, only exist to finite extent in the bulk, between the boundary $z=0$ and some cutoff surface $z_{m}$, where they `end' (meaning that one of their internal cycles shrinks down to zero size \cite{Karch:2002sh}). While this description provides a useful picture for us, we will be largely agnostic to the precise details of the flavor branes. Rather, we will simply introduce a cutoff surface at $z=z_{m}$ where the Maxwell gauge field or antisymmetric tensor field live.

A few additional comments are in order. First, it is natural to understand $z_{m}$ as a fundamental UV scale in the theory. This follows from the usual UV/IR connection in AdS/CFT, where the bulk coordinate $z$ maps to a length scale $L\sim z$ in the boundary theory such that $z=z_{m}$ corresponds to a length $L\approx z_{m}$. Second, for $z_{m}\neq0$, the fundamental matter degrees of freedom added at the boundary acquire a finite mass (or energy). This is easily understood as the introduction of this UV scale removes high energy modes, which would provide any excitation with infinite energy. Third, the existence of such a scale provides a non-zero thickness to the membranes analyzed here. The thickness is linked to the aforementioned UV/IR connection, and the fact that a finite $z_m$ introduces a minimal length one can resolve in the theory. Lastly, since $z=z_{m}\neq0$ is not exactly on the boundary, non-normalizable modes of the bulk fields, including the metric, are allowed to fluctuate on this surface. Thus, even though in our setup we are assuming a rigid boundary metric, one may couple field theoretic degrees of freedom (\emph{e.g.}, those resulting from integrating out the UV) to dynamical gravity, reminiscent of Randall-Sundrum braneworld models \cite{Randall:1999vf,Randall:1999ee}. In such a setting one would need to add a codimension-1 brane at the cutoff surface $z=z_m$ with induced dynamical gravity and proper field theoretic degrees of freedom which may backreact on the classical geometry in a consistent way (cf. \cite{Emparan:2002px,Emparan:2020znc,Emparan:2022ijy}).

\subsection{Schwinger effect: Euclidean instanton, decay rate and critical field}

\begin{figure}[t!]
\centering
\includegraphics[width=2.4in]{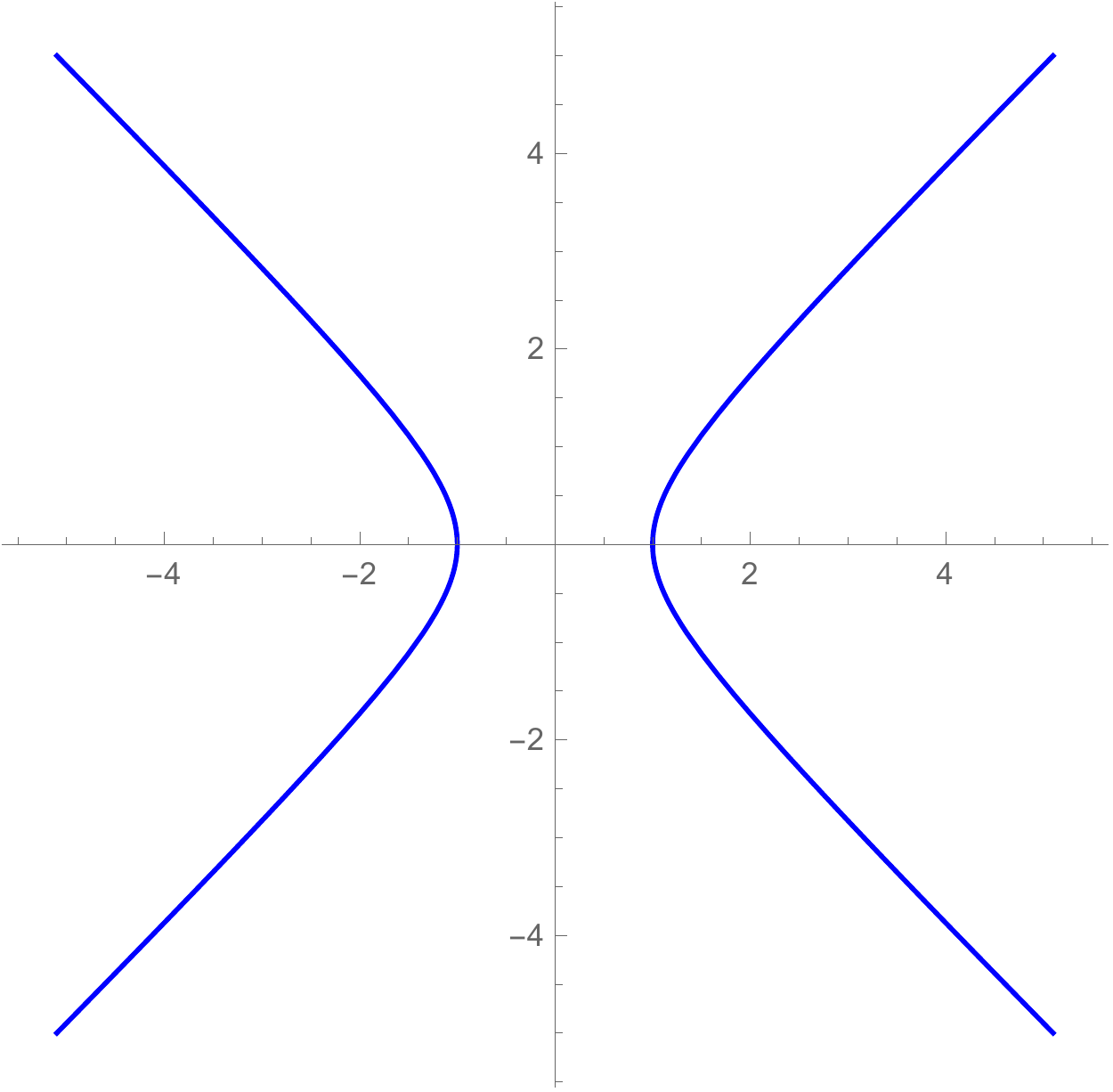}$\qquad\quad$\includegraphics[width=2.34in]{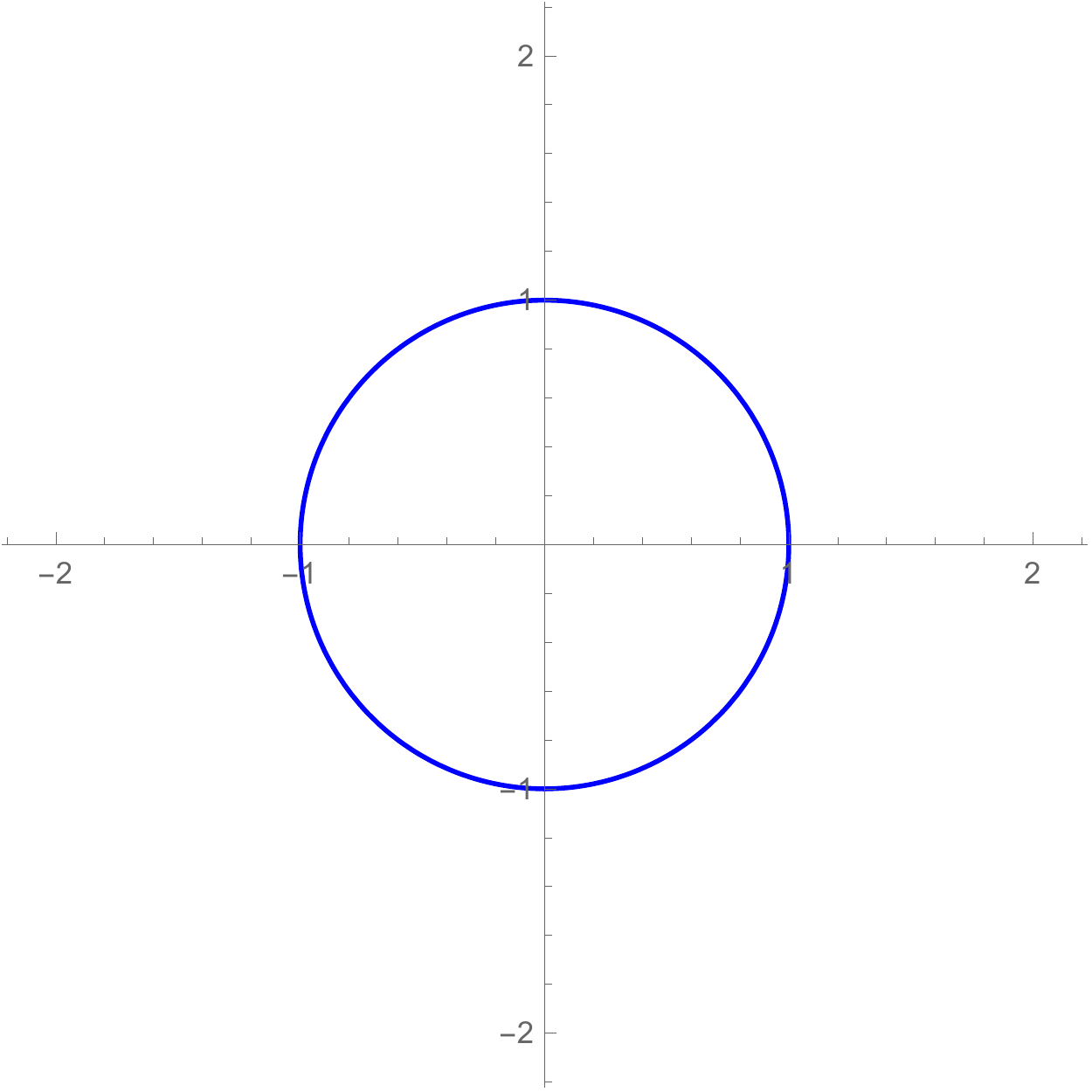}
 \put(-285,163){\small $t$}
 \put(-81,162){\small $t_E$}
 \put(-211,88){\small $x$}
 \put(-8,88){\small $x$}
 \caption{Left: Hyperbolic trajectory of a pair of particles undergoing constant proper acceleration. Right: Euclidean analytic continuation of the instanton describing pair creation. \label{figParticles}}
\end{figure}

As a warmup before we analyze membrane nucleation, let us consider a string moving in one of the spatial directions of the flat $\text{AdS}$ background (\ref{eq:flatads2back}), say $x_1\equiv x$. In holography, the string is dual to a Schwinger pair of particles. The string is attached to a stack of flavor branes that lie at $z=z_m$. Working in static gauge $\xi^i=(t,x)$, and parametrizing the string embedding as $X^\mu=(t,x,\vec{0},z(t,x))$, the Nambu-Goto action for the string yields
\be\label{NG}
S_{\text{NG}}=T_{\text{st}}\int d^2\xi \sqrt{-\det \gamma_{ij}}=\frac{\sqrt{\lambda}}{2\pi}\int_{\Sigma} dt dx \frac{\sqrt{1+z'^2-\dot{z}^2}}{z^2}\,,
\ee
where $\gamma_{ij}=\partial_iX^\mu\partial_jX^{\nu}g_{\mu\nu}$ is the induced metric on the string worldsheet $\Sigma$ and  $T_{\text{st}}=\frac{1}{2\pi\alpha'}=\frac{\sqrt{\lambda}}{2\pi L^2}$ is the string tension, given in terms of the 't Hooft coupling $\lambda$. One can easily verify that the following is a solution of the equations of motion \cite{Xiao:2008nr}:
\be\label{stringXiao}
z(t,x)=\sqrt{R^2+t^2-x^2}\,.
\ee
Notice this solution reaches out to the AdS boundary $z=0$. Indeed, evaluating the embedding at $z=0$ we can obtain the trajectory of the (as we will see, infinitely massive) particles dual to the string,
\be\label{hyper}
x(t)=\pm \sqrt{R^2+t^2}\,,
\ee
i.e., the usual hyperboloid trajectory for motion with constant proper acceleration $A=1/R$. Such a trajectory can be obtained by, e.g., exerting a constant electric field upon the particles.
In Euclidean signature, the trajectory of the pair must have
rotational symmetry given that the electric field acts as a magnetic field after Wick rotation. Hence, the
instanton follows the usual cyclotron orbit,
\be\label{hyperE}
x(t_E)=\pm \sqrt{R^2-t_E^2}\,.
\ee
See Figure \ref{figParticles} for an illustration.

The above string solution reaches out to the AdS boundary $z\to0$, hence the dual particles are infinitely massive. However, in order to properly analyze the Schwinger effect one needs to consider finite mass, otherwise pair production would be suppressed. This can be achieved by letting the flavor branes reach up to some finite radial distance $z_m\neq0$. The string solution in this case is the same as (\ref{stringXiao}) but truncated at some $z=z_m$. It can be written as follows \cite{Caceres:2010rm}:
\be
z(t,x)=\sqrt{z_m^2+R^2+t^2-x^2}\;,
\label{eq:zsolstring}
\ee
such that we recover (\ref{hyper}) by evaluating (\ref{eq:zsolstring}) at $z=z_m$.

Note that this is still the same solution, however, we have shifted $R^{2}\to R^{2}+z_{m}^{2}$. The cutoff parameter $z_m$ is a fundamental constant of the theory and is set by the position of the flavor branes. As such, several physical observables will end up depending on this parameter. For example, one can compute the (spacetime) momentum densities for a fundamental string
\be
\Pi_\mu=\frac{\partial \mathcal{L}_{\text{NG}}}{\partial \dot{X}^\mu}=-T_{\text{st}}\frac{\dot{X}_\mu X'^2-X'_\mu(\dot{X}\cdot X')}{\sqrt{(\dot{X}\cdot X')^2-\dot{X}^2X'^2}}\,,
\label{eq:momdensstring}\ee
and write them in terms of $z_m$. The prototypical example is a static string hanging from the stack of flavor branes at some $x=$ constant. In this case, the total energy $\mathcal{E}$ yields the rest mass $m$ of the dual particle,
\be
\mathcal{E}=\int_{z_m}^\infty \Pi_t dz=T_{\text{st}}L^2\int_{z_m}^\infty\frac{dz}{z^2}=\frac{\sqrt{\lambda}}{2\pi z_m}\equiv m\,,
\label{eq:restmassstring}\ee
which can be inverted to obtain
\be\label{eq:zmquark}
z_m=\frac{\sqrt{\lambda}}{2\pi m}\,.
\ee
As advertised, the mass of the particle $m$ turns out to be inversely proportional to the cutoff radius $z_m$, so that $m\to\infty$ as $z_m\to0$. We can repeat the exercise for the accelerated solution with embedding (\ref{eq:zsolstring}). In this case we obtain
\be\label{eq:energystring}
\mathcal{E}=\int_{z_m}^{\sqrt{R^{2}+z_{m}^{2}+t^{2}}} \Pi_t dz=\frac{2T_{\text{st}}L^{2}}{\sqrt{R^{2}+z_{m}^{2}}}\frac{\sqrt{R^{2}+t^{2}}}{z_{m}}\;.
\ee
The energy varies with time as the speed of the dual particles is changing. The rest energy $\mathcal{E}_{0}\equiv\mathcal{E}(t=0)$ for this solution is
\beq \mathcal{E}_{0}=\frac{\sqrt{\lambda}}{\pi\sqrt{R^{2}+z_{m}^{2}}}\left(\frac{R}{z_{m}}\right)\;,\eeq
which can also be inverted to obtain $z_m(\mathcal{E}_0,R)$.

We note that for $z_m\ll R$, or equivalently, $\sqrt{\lambda}\ll R\mathcal{E}_{0}$, then $z_{m}\approx \frac{\sqrt{\lambda}}{\pi \mathcal{E}_{0}}$, where we may relate the rest energy to the mass of a pair of static particles, $\mathcal{E}_{0}\approx2m$, recovering (\ref{eq:zmquark}). A couple of comments are in order. First note that, as in the static case, the rest energy $\mathcal{E}_{0}$ is infinite in the limit $z_{m}\to0$. This is due to the infinite volume of AdS near the conformal boundary. Second, in general $\mathcal{E}_{0}\neq2m$, rather, $\mathcal{E}_{0}$ ends up depending on $R=1/A$, which is a parameter of the specific trajectory. This is due to the fact that, for $z_m\neq0$, the dual particles are no longer pointlike, but rather, are surrounded by a `glue cloud' of finite size $z_m$ \cite{Hovdebo:2005hm,Chernicoff:2009re,Chernicoff:2009ff}. The effects of such a cloud are most directly seen from expectation values of local operators, such as $\langle \text{Tr}\, F^2(x)\rangle$ or $\langle T_{\mu\nu}(x)\rangle$, which for accelerated trajectories end up depending on not only the relativistic boost factor $\gamma(t)$, but also, the acceleration $a(t)$ and higher order derivatives, $j(t)\equiv\dot{a}(t)$, \emph{etc}, rendering these observables anisotropic \cite{Chernicoff:2011vn,Agon:2014rda}. For such extended objects, thence, the total intrinsic energy $\mathcal{E}$ (as well as their rate of radiation) ends up depending highly non-trivially on the trajectory $x(t)$, even when they are instantaneously at rest.\footnote{See \cite{Chernicoff:2011xv} for a review on the subject.}

For all practical purposes, then, we may consider $z_m$ as a fundamental constant of the theory which we will take as given. This parameter can be thought of as a UV scale giving rise to the (finite) masses/energies of the various objects that the dual theory may nucleate.

Now, to enforce the accelerated motion one must include an external force which acts upon the endpoints of the string. This can be achieved by
turning on a constant electric field with strength $F_{tx}=E$ on the flavor branes. This amounts to adding to the Nambu-Goto action (\ref{NG}) the action $S_{A}$ of a minimally coupled gauge field $A$ obeying $F=dA$,
\be\label{Gauge}
S_A=\int_{\partial\Sigma}A\,,\qquad A\equiv A_i dx^i\,,
\ee
with $\partial\Sigma$ being the boundary of the string worldsheet living on the surface $z=z_{m}$.

It turns out this system, described by the total action $S=S_{\text{NG}}+S_{A}$, is unstable above a certain critical value of the electric field strength. This can be seen from the nucleation rate of the string $\Gamma$, dual to the nucleation rate of Schwinger pair production. Indeed, from the original computation of pair creation by Schwinger \cite{Schwinger:1951nm}, one can see there is a critical value of the electric field $E_{c}=m^{2}/\alpha_{s}$, with fine structure constant $\alpha_{s}\simeq1/137$, at which point the vacuum becomes unstable and decay. Further, since the critical field strength is beyond the limit of weak field condition, $E_{c}\gg m^{2}$, one may expect nucleation processes to receive relevant non-perturbative corrections.
Motivated by this, the Schwinger pair production rate (\ref{eq:Schwingerrate}) may be derived rederived using instanton techniques, specifically, from
the imaginary part of the Euclidean worldline path integral, (see e.g. \cite{Cohen:2008wz}), where the contribution of an instanton to the nucleation rate is of the form $\Gamma=A e^{-S_E}$, where $A$ is a prefactor that comes from integrating over quantum fluctuations about the instanton solution and $S_E$ is the on-shell Euclidean  action.

\begin{figure}[t!]
\centering
\includegraphics[width=2.4in]{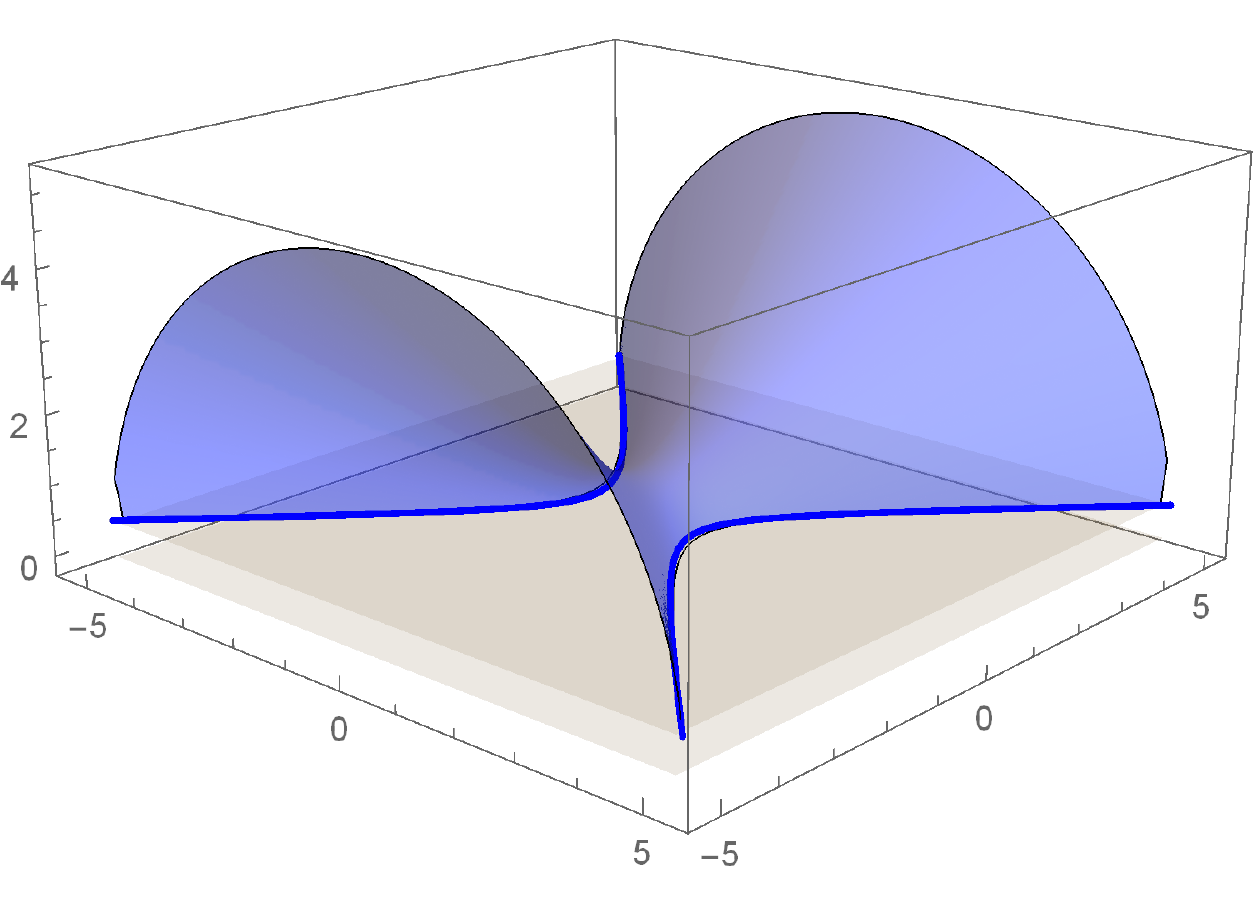}$\qquad\quad$\includegraphics[width=2.4in]{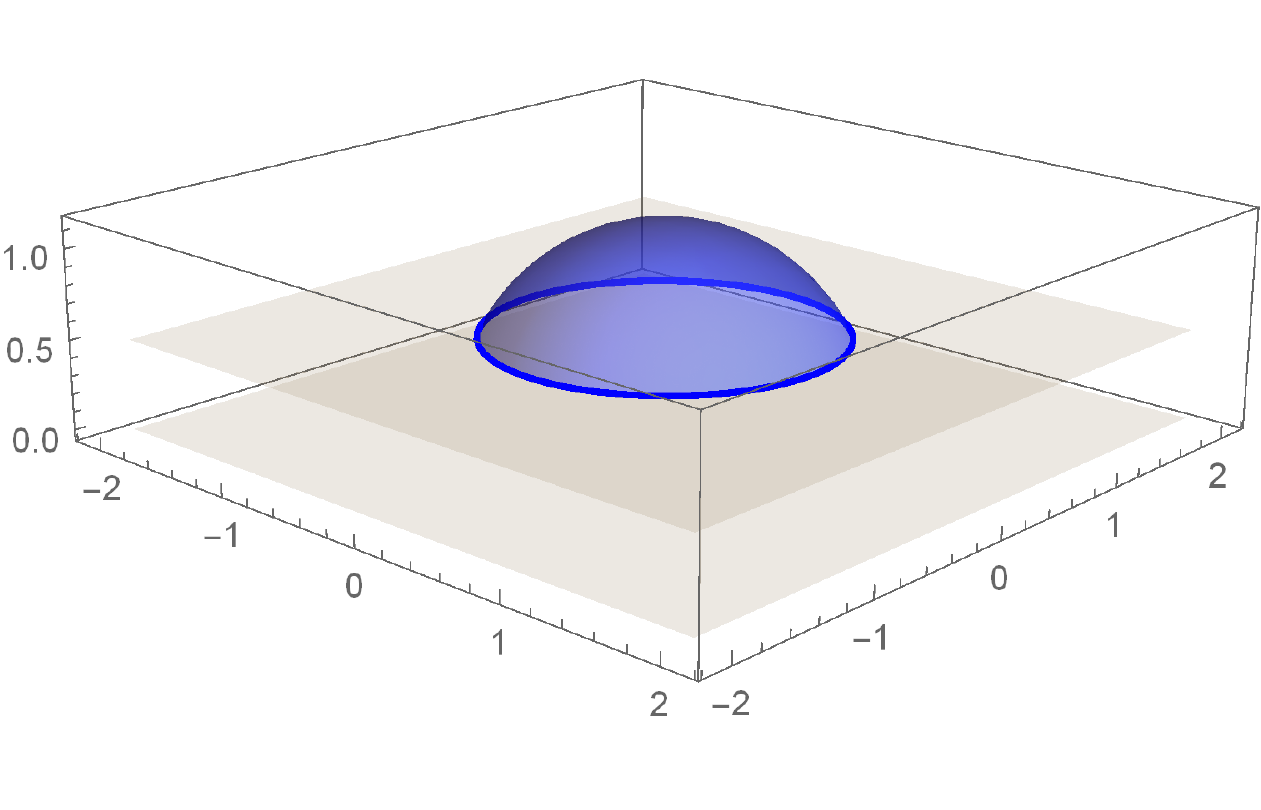}
 \put(-241,20){\small $t$}
 \put(-342,20){\small $x$}
  \put(-32,22){\small $t_E$}
 \put(-132,22){\small $x$}
 \put(-387,72){\small $z$}
 \put(-182,62){\small $z$}
 \put(-280,45){\small $z=z_m$}
 \put(-84,49){\small $z=z_m$}
 \caption{Left: Lorentzian string solution dual to a pair of particles undergoing constant proper acceleration. Right: String solution dual to the Euclidean instanton. In both cases the solutions are truncated at some $z=z_m$ where the embeddings coincide with the boundary Lorentzian and Euclidean trajectories, (\ref{hyper}) and (\ref{hyperE}), respectively. \label{figStrings}}
\end{figure}
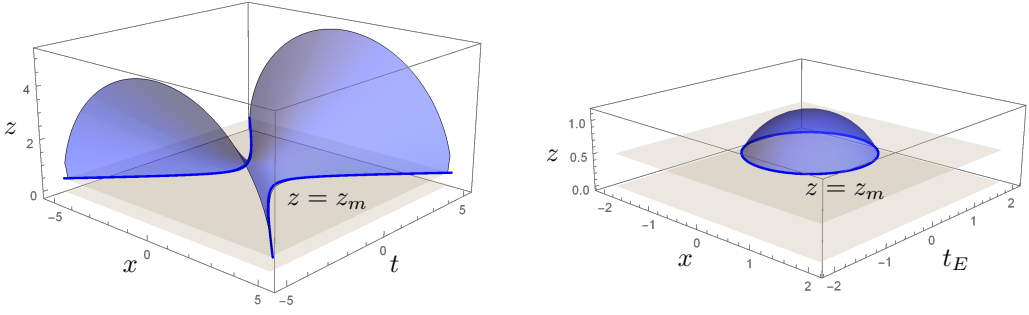

Missing from this derivation, however, are effects coming from the backreaction of the particles on the gauge fields, which are enhanced at strong coupling. To study such corrections, one may use AdS/CFT to reexamine the Schwinger effect, holographically understood as the nucleation of the string described above \cite{Semenoff:2011ng}. Thus, one moves to Euclidean signature, $t\to -i t_{E}$ for Euclidean time $t_{E}$, and looks for instanton solutions to the Euclideanized Nambu-Goto action. In this case the instanton is given by the Wick rotated string solution (\ref{eq:zsolstring}), which takes the following form:
\be\label{eq:zsolstringE}
z(t_E,x)=\sqrt{z_m^2+R^2-t_{E}^2-x^2}\,.
\ee
Figure \ref{figStrings} shows a schematic illustration of the Lorentzian and Euclidean string embeddings, (\ref{eq:zsolstring}) and (\ref{eq:zsolstringE}), respectively.

Note that at $z=z_m$, the Euclidean solution reduces to a circle of radius $R$, as expected from (\ref{hyperE}). Hence,  an  instanton  of  the Euclidean path integral is a cyclotron orbit which may wrap $n$ times in the time direction.
The associated Euclidean action, $S_E=S_{\text{NG}}+S_{A}|_{t\to -i t_{E}}$, evaluated on-shell is found to be
\be\label{SEstring}
S_E=n \sqrt{\lambda}\left(\sqrt{1 +\frac{R^2}{z_m^2}}-1\right)-n\pi R^2E\,.\\
\ee
For $z_m\ll R$ (i.e., $\sqrt{\lambda}\ll R\mathcal{E}_{0}$), $z_m\approx \frac{\sqrt{\lambda}}{\pi\mathcal{E}_0}\approx \frac{\sqrt{\lambda}}{2\pi m}$, the action reproduces the weak coupling result of \cite{Cohen:2008wz}, however, corrected by a term $-n\sqrt{\lambda}$, representing the inclusion of a Wilson loop amplitude
in the path integral \cite{Semenoff:2011ng}. This extra term encapsulates the effects of strong coupling which, as advertised above, we have a handle on only through the power of holography.

The radius $R$ of the orbit can be fixed by finding an extremum of (\ref{SEstring}),
\be\label{eq:solR}
R=\sqrt{\frac{\lambda}{4\pi^2E^2z_m^2}-z_m^2}\,.
\ee
Consequently, this yields the following critical value $E_{c}$ of the field strength (when $R\to0$, or equivalently, $A\to\infty$),
\be\label{Ecstring}
E_c=\frac{\sqrt{\lambda}}{2\pi z_m^2}\,,
\ee
which may be used to rewrite (\ref{eq:solR}) as
\be
R=\sqrt{\frac{\sqrt{\lambda}}{2\pi E_c}\left(\frac{E_c^2}{E^2}-1\right)}
\ee
Similarly, the on-shell action (\ref{SEstring}) may be cast in terms of $E_c$ as
\be
S_E=\frac{n\sqrt{\lambda}}{2}\frac{(E_c-E)^2}{E_cE}\,.
\ee
Since $S_E$ vanishes as $E\to E_c$, one sees that
when this critical field is reached, the instanton sum is
no longer exponentially suppressed, rendering
the vacuum unstable.   Moreover, it is easy to check that in the weak field limit $E \ll E_c$, and $z_m\ll R$ (or $\sqrt{\lambda}\ll R\mathcal{E}_{0}$), one recovers the argument in the exponential of Schwinger's result (\ref{eq:Schwingerrate}), valid at weak coupling.

As a final comment, we note that the critical field strength $E_{c}$ can be easily understood from the flavor brane point of view. Specifically, in the static gauge, the DBI Lagrangian for a probe brane reads
\be
\sqrt{-\det(g_{ab}+2\pi\alpha' F_{ab})}\propto\sqrt{\frac{1}{z^4}-\left(\frac{2\pi}{\sqrt{\lambda}}F_{tx}\right)^2}\,,
\ee
which is real at $z=z_m$ provided $F_{tx}\equiv E<\sqrt{\lambda}/(2\pi z_m^2)$. Beyond this value, the creation of open strings is energetically favored, such that the system becomes unstable, as $E>E_{c}$.

%%%%%%%%%%%%%%%%%%%%%%%%%%%%%%%%%%%%%%%%%%%%%%%%%%%%%%%%%%

\section{Membrane nucleation rates in flat space} \label{sec:flatrates}

Having reviewed the nucleation of strings in flat space, we generalize to the case of nucleating spherical membranes, first in ($d+1$)-dimensional flat space, and then in ($d+1$)-dimensional de Sitter space in Section \ref{sec:dSrates}.

\subsection{Membranes in flat space}

Let us consider a $(p+1)$-dimensional membrane embedded in pure $\text{AdS}_{d+2}$, with $1\leq p\leq d$, where in the limit $p=1$ the system reduces to a string as reviewed in Section \ref{sec:setupreview}. Assuming spherical polar coordinates along the $p$-spatial directions, the bulk $\text{AdS}_{d+2}$ metric in Poincar\'e coordinates is
\beq ds_{d+2}^{2}=\frac{L^{2}}{z^{2}}(-dt^{2}+dr^{2}+r^{2}d\Omega_{p-1}^{2}+d\vec{x}_{\perp}^{2}+dz^{2})\;.\label{eq:AdSmetPoinc}\eeq
Here $\vec{x}_{\perp}$ denotes the directions transverse to the brane; however, when $p=d$, there are no extra transverse directions. As before, we imagine flavor branes extending into the bulk from the boundary at $z=0$ to some finite cutoff position $z=z_{m}$. The spherical membranes are attached to the stack of flavor branes along $z=z_{m}$, such that the boundary of the membrane worldvolume ends at $z=z_{m}$. In a moment we will see how to relate the energy of various membrane solutions the bulk cutoff position $z_{m}$, analogous to the string case.

Working in static gauge, the worldvolume $\Sigma_{p}$ is parametrized  by coordinates $\xi^{i}=(t,r,\vec{\Omega}_{p-1})$ while the embedding functions of the membrane into $\text{AdS}_{d+2}$ are $X^{\mu}=(t,r,\vec{\Omega}_{p-1},\vec{x}_{\perp},z(t,r))$. Here $\vec{\Omega}_{p-1}$ denotes the collection of spherical angular coordinates, and $\vec{x}_{\perp}$ denotes the coordinates of the $(p-d)$ transverse directions, which, without loss of generality, we may fix to be at $\vec{0}$. The Lorentzian  action $S_{p}$ characterizing the dynamics of the membrane is given by the area of $\Sigma_{p}$, generalizing the Nambu-Goto action for a string:
\beq S_{p}=T_{p}\int_{\Sigma_{p}}d^{p+1}\xi\sqrt{-\text{det}\gamma_{ij}}\;.\label{eq:NGactgenp}\eeq
Here $T_{p}$ denotes the tension of the membrane, such that for $p=1$ we recover the tension of the string $T_{1}=\frac{\sqrt{\lambda}}{2\pi L^{2}}$. With respect to the line element (\ref{eq:AdSmetPoinc}), it is straightforward to show the action becomes
\beq S_{p}=T_{p}L^{p+1}\Omega_{p-1}\int_{\Sigma_{p}} dtdr\frac{r^{p-1}\sqrt{1+z'^{2}-\dot{z}^{2}}}{z^{p+1}}\;,\label{eq:NGpflat}\eeq
where $\Omega_{p-1}=\frac{2\pi^{p/2}}{\Gamma(p/2)}$ is the volume of a $(p-1)$-dimensional unit sphere, and we have introduced the shorthand notation $\dot{z}\equiv\frac{dz}{dt}$  and $z'\equiv\frac{dz}{dr}$.

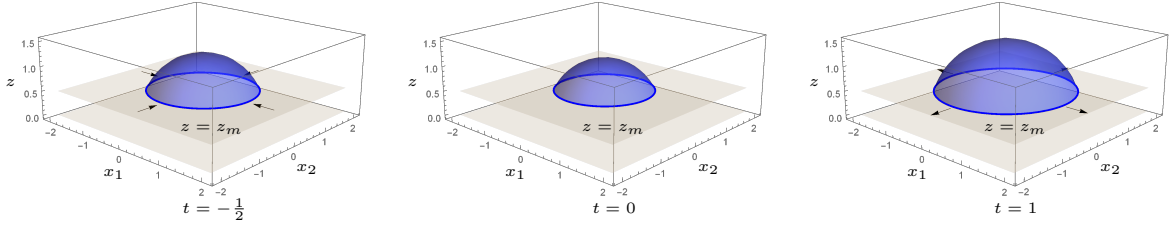
\begin{figure}[t!]
\centering
\includegraphics[width=1.8in]{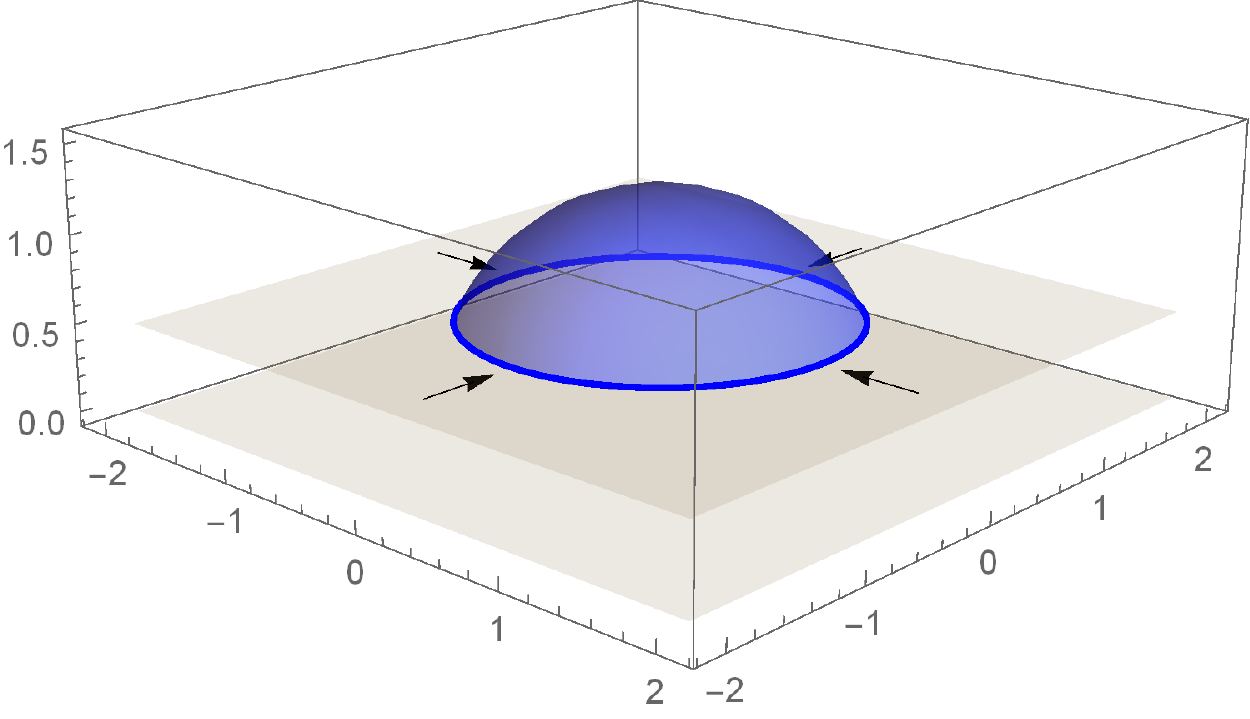}$\qquad$\includegraphics[width=1.8in]{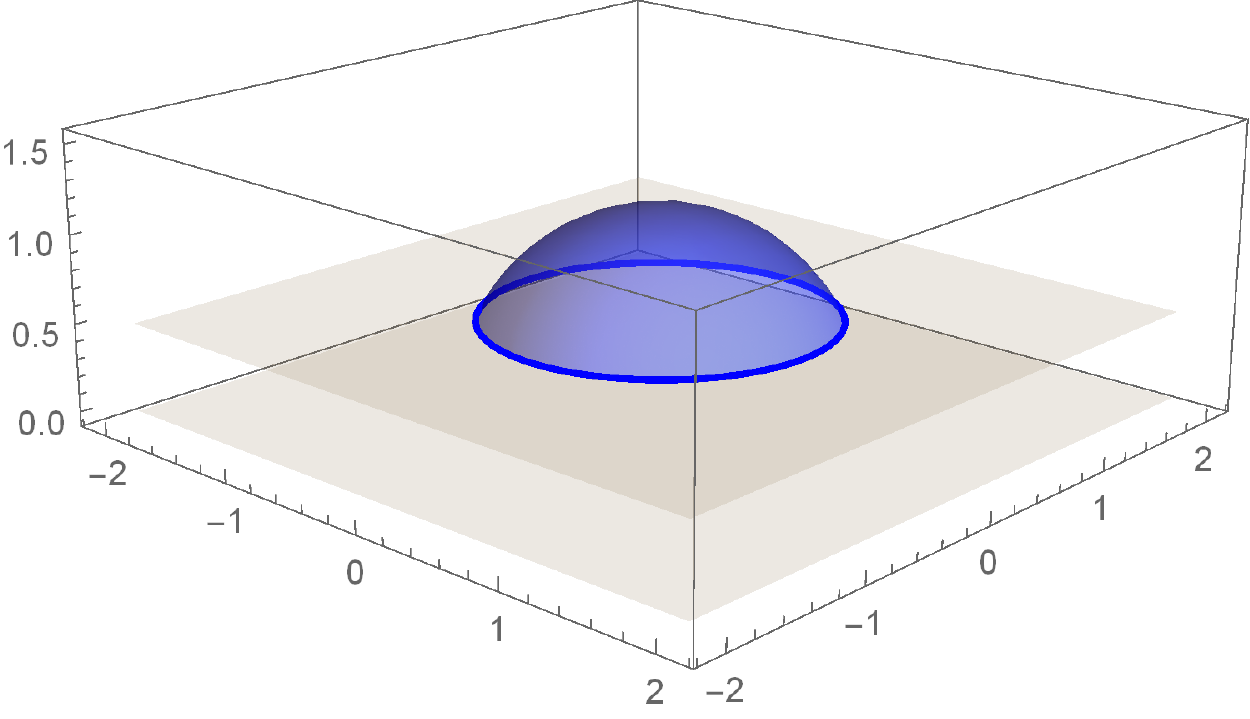}$\qquad$\includegraphics[width=1.8in]{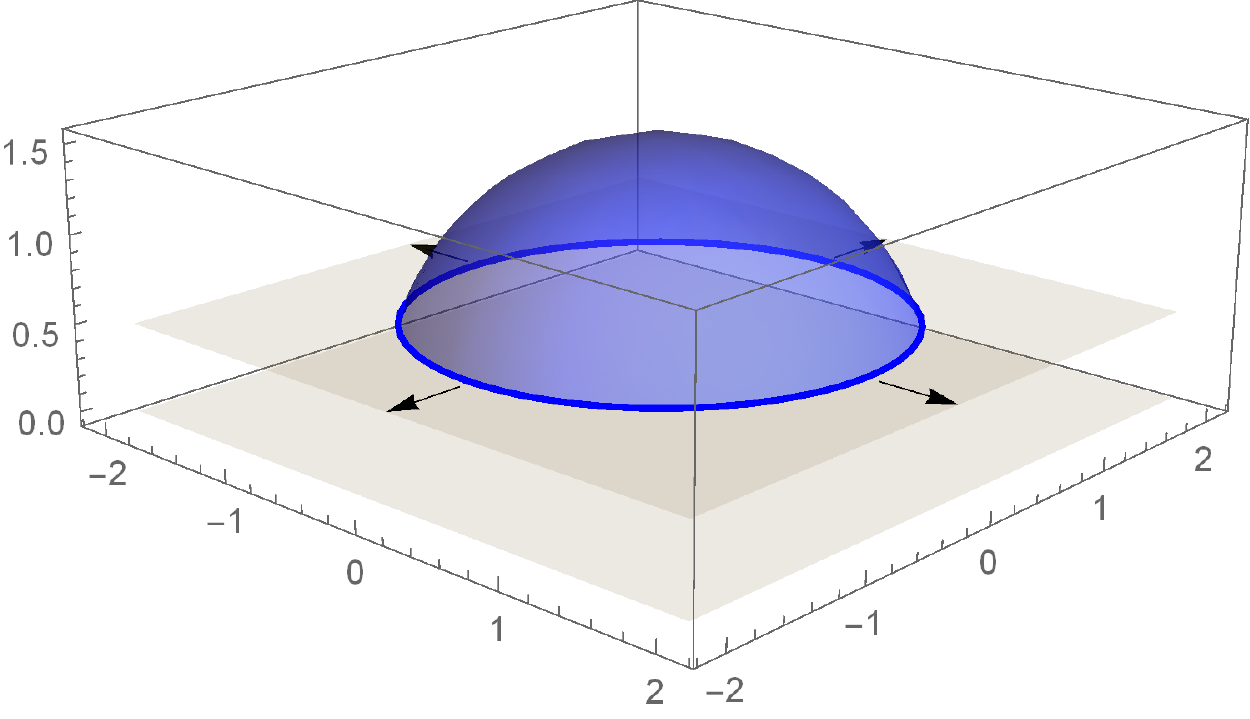}
 \put(-403,9){\tiny $x_1$}
\put(-251,9){\tiny $x_1$}
\put(-99,9){\tiny $x_1$}
 \put(-330,11){\tiny $x_2$}
\put(-178,11){\tiny $x_2$}
\put(-26,11){\tiny $x_2$}
 \put(-440,43){\tiny $z$}
 \put(-288,43){\tiny $z$}
 \put(-136,43){\tiny $z$}
 \put(-70,26){\tiny $z=z_m$}
\put(-222,26){\tiny $z=z_m$}
\put(-374,26){\tiny $z=z_m$}
 \put(-66,-5){\tiny $t=1$}
\put(-218,-5){\tiny $t=0$}
\put(-373,-5){\tiny $t=-\frac{1}{2}$}
 \caption{Snapshots of the (Lorentzian) membrane solution (\ref{eq:zsolnflatmem}) as it contracts and expands in time. For the sake of the plots, we have only shown the case of a $(2+1)$-dimensional membrane (\emph{i.e.}, with $p=2$), but the solution is valid for arbitrary $p$. The $p=1$ case reduces to a string worldsheet, as depicted in Figure \ref{figStrings} (left).
\label{figMem}}
\end{figure}

From the action (\ref{eq:NGpflat}), one can easily verify the same solution $z(t,r)$ for the string (\ref{eq:zsolstring}) is a solution to the equations of motion of the membrane for any $p$
\beq z(t,r)=\sqrt{R^{2}+z_{m}^{2}+t^{2}-r^{2}}\;,\label{eq:zsolnflatmem}\eeq
with the change $x\leftrightarrow r$ and where we have explicitly included the cutoff $z_{m}$ such that the membrane has finite mass (energy). Thus, at the cutoff surface $z=z_{m}$, the solution (\ref{eq:zsolnflatmem}) describes a massive spherical membrane which contracts and expands radially with radius given by $r(t)=\sqrt{R^{2}+t^{2}}$. Figure \ref{figMem} illustrates the time evolution of this membrane solution.

As for the case of the holographic Schwinger effect, enforcing this solution requires an external force acting on the boundary of the membrane $\partial\Sigma_{p}$. Unlike the string case, however, this amounts to coupling the membrane to a higher antisymmetric tensor field $A$ of rank $p$,
\beq A=A_{[i...k]}dx^{i}\wedge...\wedge dx^{k}\;,\eeq
on the boundary where one imagines flavor branes to extend. The action for $A$ is
\beq S_{A}=\int_{\partial\Sigma_{p}}\hspace{-3mm}A\;.\eeq
For spherically symmetric configurations, the gauge field strength $F$ associated with $A$ via $F=dA$, will only have a single independent component, and may be expressed as
\beq F=-E\tilde{\epsilon}\;,\eeq
where $\tilde{\epsilon}$ is the natural volume form of the ($p+1$)-dimensional spacetime, $\tilde{\epsilon}=\sqrt{|g|}\epsilon_{i\ldots kl}dx^i\wedge\ldots\wedge dx^k\wedge dx^l$ (with $\epsilon_{01\ldots p}= 1$),
and $E$ is a constant `electric' field strength. The total action of interest is then
\beq S=S_{p}+S_{A}\;.\label{eq:totLoractflat}\eeq

\subsubsection*{Rest energy of membranes}

We can compute the energy $\mathcal{E}$ of any membrane solution, analogous to the string case (\ref{eq:restmassstring}) or (\ref{eq:energystring}), and determine its dependence on the various parameters of the theory.

It is first worthwhile to consider the energy of a static membrane that hangs at some $x_1=$ constant and extends to infinity in the remaining $p-1$ directions. In this case, the total energy $\mathcal{E}$ yields the rest mass $m_p$ of the membrane, generalizing (\ref{eq:restmassstring}) to:
\be
\mathcal{E}=\int_{z_m}^\infty \Pi_t dz=T_{p}L^{p+1}V_{p-1}\int_{z_m}^\infty\frac{dz}{z^{p+1}}=\frac{ T_p\,L^{p+1}V_{p-1}}{p\, z_m^p}\equiv m_p\,.
\label{eq:restmassmem}\ee
Here $V_{p-1}\equiv\int dx_2\cdots dx_p$ stands the $p-1$ volume. If desired, we can invert (\ref{eq:restmassmem}) to express $z_m$ in terms of $m_p$, which yields
\be\label{zmanyp}
z_m=\left(\frac{T_p\,L^{p+1}V_{p-1}}{p\,m_p}\right)^{1/p}\,,
\ee
which correctly reproduces (\ref{eq:zmquark}) for $p=1$ with $T_1\to\frac{\sqrt{\lambda}}{2\pi L^2}$ and $V_{p-1}\to1$.

We can now consider the spherical membranes with profile given by (\ref{eq:zsolnflatmem}). In this case, the momentum densities for a given spherically symmetric solution are given by:
\be
\Pi_\mu=\frac{\partial \mathcal{L}_p}{\partial \dot{X}^\mu}=-T_{p}\Omega_{p-1}\left(\frac{Lr}{z}\right)^{p-1}\left(\frac{\dot{X}_\mu X'^2-X'_{\mu}(\dot{X}\cdot X')}{\sqrt{(\dot{X}\cdot X')^2-\dot{X}^2X'^2}}\right)\;.
\ee
Substituting in the solution (\ref{eq:zsolnflatmem}), the energy of the membrane yields,\footnote{To carry out this calculation it is easier to parametrize the worldvolume with coordinates $\xi^{i}=(t,z,\vec{\Omega}_{p-1})$ and invert (\ref{eq:zsolnflatmem}) to obtain $r(t,z)$, such that we work with the embedding $X^{\mu}=(t,r(t,z),\vec{\Omega}_{p-1},\vec{x}_{\perp},z)$.}
\beq \mathcal{E}=\int_{z_{m}}^{\sqrt{R^{2}+t^{2}+z_{m}^{2}}}dz\Pi_{t}=\frac{T_{p}\Omega_{p-1}L^{p+1}}{p\sqrt{R^{2}+z_{m}^{2}}}\frac{(R^{2}+t^{2})^{p/2}}{z_{m}^{p}}\;.\eeq
As expected, the energy decreases as the membrane contracts, reaches a minimum at $t=0$ and then grows back to infinity as the membrane expands. The rest energy of the membrane $\mathcal{E}_{0}\equiv\mathcal{E}(t=0)$ is thus
\beq \mathcal{E}_{0}=\frac{T_{p}\Omega_{p-1}L^{p+1}}{p\sqrt{R^{2}+z_{m}^{2}}}\left(\frac{R}{z_{m}}\right)^{p}\;.\label{eq:E0flatmemexact}\eeq
As before, we assume $z_m$ to be a fundamental constant of the theory, which we take as given. One may try to invert (\ref{eq:E0flatmemexact}) to express the cutoff $z_m$ in terms of $\mathcal{E}_{0}$ and $R$, but the result is lengthy and not particularly useful, hence we will refrain from doing so.

As a final comment we point out that, as for the case of the string, the truncated solution with $z_m\neq0$ is expected to be dual to a membrane with a finite width proportional to $z_m$, thus departing from the infinitely thin shell regime. It would be interesting to investigate this further. For example, it would be insightful to compute the expectation values of local operators, such as $\langle \text{Tr}\, F^2(x)\rangle$ or $\langle T_{\mu\nu}(x)\rangle$, \emph{e.g.}, following  \cite{Chernicoff:2011vn,Agon:2014rda}, and determine the spatial profiles sourced by these membranes.

.

\subsection{On-shell Euclidean action and nucleation rate}

The nucleation rate $\Gamma$ for membranes, analogous to Schwinger pair production, can be computed using instanton methods \cite{Garriga:1993fh}, where the leading order effect is given by the on-shell Euclidean action $S_{E}$ (\ref{eq:nucrategenintro}). Thus far, aside from the case of the string \cite{Fischler:2014ama}, such nucleation rates have been carried out at weak coupling, without the employ of holography. Here we use AdS/CFT as a general tool to compute $\Gamma\sim e^{-S_{E}}$, for which the classical bulk computation is valid, leaving aside the corrections due to quantum fluctuations. We thus Euclideanize the total action (\ref{eq:totLoractflat}) via Wick rotating the $\text{AdS}_{d+2}$ time $t\to -i t_{E}$,
\beq S_{E}=S^{E}_{p}+S^{E}_{A}\;,\label{eq:totalactEuc}\eeq
and use the fact that the field strength tensor $F$ is unchanged under the Wick rotation (as is the case for standard Maxwell theory).

Specifically, Euclideanizing the membrane action (\ref{eq:NGpflat}) leads to
\beq S_{p}^{E}=T_{p}L^{p+1}\Omega_{p-1}\int_{\Sigma_{p}} dt_{E}dr\frac{r^{p-1}\sqrt{1+z'^{2}+\dot{z}^{2}}}{z^{p+1}}\;,\label{eq:EucactNGflat}\eeq
where now $\Sigma_{p}$ is the Euclidean worldvolume and $\dot{z}\equiv \frac{\partial z}{\partial t_{E}}$. The solution (\ref{eq:zsolnflatmem}) Wick rotates to
\beq z(t_{E},r)=\sqrt{R^{2}+z_{m}^{2}-t_{E}^{2}-r^{2}}\;,\eeq
and is a solution to the Euclidean equations of motion coming from (\ref{eq:EucactNGflat}). It corresponds to a spherical membrane, where $t_E$ now acts as an additional spatial coordinate, and can be viewed as a higher dimensional generalization to the Euclidean string solution shown in Figure \ref{figStrings} (right). In this case, the on-shell Euclidean action for the membrane yields
\beq S^{E}_{p}=T_{p}L^{p+1}\Omega_{p-1}\sqrt{R^{2}+z_{m}^{2}}\int_{\Sigma_{p}} dt_E dr\frac{r^{p-1}}{[R^{2}+z_{m}^{2}-t_E^{2}-r^{2}]^{(p+2)/2}}\;.\eeq
To evaluate the integral we perform the following change of integration variables,
\beq r=\rho\cos\phi\;,\quad t_{E}=\rho\sin\phi\;,\eeq
where since $r$ is taken to be non-negative, $\rho\in[0,R]$ and $\phi\in[-\frac{\pi}{2},\frac{\pi}{2}]$, in contrast with the string case, where $x$ ranges over positive and negative values. Consequently, the on-shell membrane action evaluates to
\be\label{SpE}
S_p^E=\frac{\pi ^{\frac{p+1}{2}}   T_p \,L^{p+1}R^{p+1}}{\Gamma(\frac{p+3}{2})(z_m^2+R^2)^{\frac{p+1}{2}}}\, _2F_1\left(\frac{p+1}{2},\frac{p+2}{2},\frac{p+3}{2},\frac{R^2}{R^2+z_m^2}\right)\,.
\ee
where we have used $\Omega_{p-1}=\frac{2\pi^{p/2}}{\Gamma(p/2)}$. For later convenience, we also use the following identity of hypergeometric functions:
\be
z_m^p\,\!_2F_1\left(\frac{p+1}{2},\frac{p+2}{2},\frac{p+3}{2},\frac{R^2}{R^2+z_m^2}\right)=(R^2+z_m^2)^{\frac{p}{2}}\,\!_2F_1\left(\frac{1}{2},1,\frac{p+3}{2},\frac{R^2}{R^2+z_m^2}\right)\,,
\ee
to rewrite (\ref{SpE}) as
\beq S_{p}^{E}=\frac{\pi^{\frac{p+1}{2}}T_{p}\,L^{p+1}R^{p+1}}{\Gamma(\frac{p+3}{2})z_{m}^{p}\sqrt{R^{2}+z_{m}^{2}}} {}_2F_{1}\left(\frac{1}{2}\,,\,1\,,\,\frac{p+3}{2}\,,\,\frac{R^{2}}{R^{2}+z_{m}^{2}}\right)\;.\label{eq:SpNGflatonshellEuc}\eeq

Meanwhile, since the boundary of the Euclidean worldvolume $\Sigma_{p}$ is closed, we may employ Stokes' theorem such that, on-shell, the  second part of the action, $S^{E}_{A}$, may  be cast as
\beq S_{A}^{E}=-E\int_{\mathcal{V}_{p+1}}\tilde{\epsilon}=-E\mathcal{V}_{p+1}\;,\label{eq:SAEucflat}\eeq
 where $\mathcal{V}_{p+1}$ denotes the Euclidean spacetime volume enclosed by the boundary of the membrane, $\partial \Sigma_p$, a $(p+1)$-dimensional ball of radius $R$,
\beq \mathcal{V}_{p+1}=\frac{\pi^{\frac{p+1}{2}}R^{p+1}}{\Gamma(\frac{p+3}{2})}=\frac{\Omega_{p}R^{p+1}}{(p+1)}\;.\eeq
Combining (\ref{eq:SpNGflatonshellEuc}) and (\ref{eq:SAEucflat}), the total on-shell Euclidean action (\ref{eq:totalactEuc}) for arbitrary $p$ yields
\beq S_E=\frac{\Omega_{p}R^{p+1}}{(p+1)}\left[\frac{T_{p}\,L^{p+1}}{z_{m}^{p}\sqrt{R^{2}+z_{m}^{2}}} {}_2F_{1}\left(\frac{1}{2}\,,\,1\,,\,\frac{p+3}{2}\,,\,\frac{R^{2}}{R^{2}+z_{m}^{2}}\right)-E\right]\;.\label{eq:totalonshellEucflat}\eeq
For example,
\beq
\begin{split}
&S_E^{(p=1)}= \frac{2\pi L^2 T_1}{z_m}\left(\sqrt{R^2 +z_m^2}-z_m\right)-\pi R^2E,\\
&S_{E}^{(p=2)}=\frac{2\pi L^{3}T_{2}}{z_m^2}\left[R\sqrt{R^{2}+z_{m}^{2}}-z_m^2\text{arctanh}\left(\frac{R}{\sqrt{R^{2}+z_{m}^{2}}}\right)\right]-\frac{4}{3}\pi R^{3}E\;,\\
&S_E^{(p=3)}=\frac{2\pi^{2}L^{4}T_{3}}{3z_{m}^{3}}\sqrt{R^{2}+z_{m}^{2}}\left[R^{2}+2z_{m}^{2}\left(\sqrt{\frac{z_{m}^{2}}{R^{2}+z_{m}^{2}}}-1\right)\right]-\frac{1}{2}\pi^{2}R^{4}E\;,
\end{split}
\label{eq:flatmemsd123act}\eeq
where for $p=1$ we recover the on-shell string action (\ref{SEstring}) (for winding $n=1$).

We can fix the radius $R$ in terms of $z_{m},T_{p}$ and $E$ by finding the extremum $dS_{E}/dR=0$,
\beq R=\sqrt{\left(\frac{T_{p}\,L^{p+1}}{Ez_{m}^{p}}\right)^{2}-z_{m}^{2}}\;.\label{eq:Rrad}\eeq
As for the string case, the entire system will become unstable for a critical value of the field strength, denoted $E_{c}$, which occurs when $A\to\infty$ or $R\to0$. Explicitly,
\beq E_{c}=T_{p}\left(\frac{L}{z_{m}}\right)^{p+1}\;,\label{eq:Ecritflat}\eeq
such that the radius $R$ is
\beq R=\frac{z_{m}}{E}\sqrt{E_{c}^{2}-E^{2}}\;,\eeq
and the on-shell action (\ref{eq:totalonshellEucflat}) becomes
\beq S_{E}=\frac{\Omega_{p}}{(p+1)}ER^{p+1}\left[{}_2F_{1}\left(\frac{1}{2}\,,\,1\,,\,\frac{p+3}{2}\,,\,\frac{E_{c}^{2}-E^{2}}{E_{c}^{2}}\right)-1\right]\;.\eeq
Since $S_E\propto R^{p+1}$ it is clear the on-shell action vanishes when the field strength approaches its critical value $E\to E_{c}$. Further, given the field strength dependence in $R$, it is easy to see that for any integer $p>0$ $S_{E}$ diverges in the limit $E\to0$. Thus, there is not a consistent limit for which membrane creation may occur in flat space without coupling to the external gauge field $A$. We will see in the next section that this need not be the case for membrane nucleation in a de Sitter background.

Lastly, in the weak field limit $E\ll E_c$, and $z_m\ll R$ (or, equivalently, $T_pL^{p+1}\ll R\mathcal{E}_0$) we have $R\approx(\frac{p\mathcal{E}_{0}}{E\Omega_{p-1}})^{1/p}$, reproducing the result found using standard field theoretic methods at weak coupling \cite{Garriga:1993fh} (see Eq. (\ref{eq:Euconshellflatapp}) in Appendix \ref{app:nucweak} for details)
\beq S_{E}\approx \frac{\Omega_{p}}{p(p+1)}ER^{p+1}=\frac{\Omega_{p}}{(p+1)}\left[\frac{p}{E}\left(\frac{\mathcal{E}_{0}}{\Omega_{p-1}}\right)^{(p+1)}\right]^{1/p}\;,\label{eq:flatcheck}\eeq
where we used ${}_{2}F_{1}(\frac{1}{2},1,\frac{p+3}{2},1)=1+1/p$. More generally, expanding (\ref{eq:totalonshellEucflat}) in the two dimensionless parameters, one finds an infinite set of corrections in powers of $E/E_c$ and $T_pL^{p+1}/R\mathcal{E}_0$, which may be regarded as non-perturbative corrections to (\ref{eq:flatcheck}) due to strong coupling and strong electric fields, and away from the infinitely thin shell regime.

%%%%%%%%%%%%%%%%%%%%%%%%%%%%%%%%%%%%%%%%%%%%%%%%%%%%%%%%%%
\section{Membrane nucleation rates in de Sitter space} \label{sec:dSrates}

Thus far we efficiently computed the nucleation rate for spherical membranes including non-perturbative effects via AdS/CFT holography, where the conformal boundary of AdS was taken to be Minkowski space. Importantly, the conformal boundary of AdS where the CFT lives need not be flat, allowing one to study, at least in principle, quantum processes in curved backgrounds. As such, here we investigate membrane nucleation rates in $(d+1)$-dimensional de Sitter space ($\text{dS}_{d+1}$) at strong coupling using the tools of AdS/CFT.

As a starting point, notice one may foliate asymptotically $\text{AdS}_{d+2}$ spaces using $\text{dS}_{d+1}$ slices such that the dual theory is a strongly-coupled QFT living on a fixed $\text{dS}_{d+1}$ background \cite{Marolf:2013ioa}. In particular, when the bulk geometry is pure $\text{AdS}_{d+2}$,
then the line element can be cast in Fefferman-Graham form as \cite{Fischler:2013fba,Fischler:2014tka,Fischler:2014ama}
\beq ds^{2}_{d+2}=\frac{L^{2}}{z^{2}}\left[f(z)^{2}ds^{2}_{\text{dS}}+dz^{2}\right]\;,\quad f(z)\equiv \left(1-\frac{H^{2}z^{2}}{4}\right)\label{eq:AdSlineeledS}\eeq
and the dual theory is a CFT.\footnote{Other studies of holographic CFTs in dS that make use of this type of foliations include \cite{Zhang:2014cga,Chu:2016uwi,Chu:2016pea,Zhang:2019vgl,Ageev:2021xjk}.} Here
$ds^{2}_{\text{dS}}$ is the $(d+1)$-dimensional line element for de Sitter space in \emph{any} coordinate system, with $H$ being the Hubble constant. This metric has an acceleration horizon at $z=2/H$. Furthermore, for any static observer there is a local notion of thermodynamics, with a temperature equal to the Gibbons-Hawking temperature of the acceleration horizon, $T_{\text{dS}}=\frac{H}{2\pi}$. It is useful to redefine the bulk $z$ coordinate by
\beq z=\frac{2}{H}e^{-2\text{arctanh}(u)}=\frac{2}{H}\frac{(1-u)}{(1+u)}\;,\label{eq:coordzr}\eeq
such that $u\in[0,1]$, with $u=0$ corresponding to the acceleration horizon ($z=2/H$) and $u=1$ the conformal boundary of AdS ($z=0$).
Thence, the line element (\ref{eq:AdSlineeledS}) becomes
\beq ds^{2}_{d+2}=\frac{4L^{2}}{(1-u^{2})^{2}}\left[H^{2}u^{2}ds_{\text{dS}}^{2}+du^{2}\right]\;.\label{eq:AdSmetdS}\eeq

In Appendix \ref{app:dSfoliationsAdS} we analyze multiple such foliations of $\text{AdS}_{d+2}$ due to various coordinate representations of $\text{dS}_{d+1}$. Two particular forms of the $\text{AdS}_{2}$ line element (\ref{eq:AdSlineeledS})  of interest include foliations by de Sitter described in global coordinates,
\beq ds^{2}_{d+2}=\frac{4L^{2}}{(1-u^{2})^{2}}\left[H^{2}u^{2}\left(-d\tau^{2}+\frac{\cosh^{2}(H\tau)}{H^{2}}d\Omega_{d}^{2}\right)+du^{2}\right]\;,\label{eq:globaldSfol}\eeq
and static patch coordinates,
\beq ds^{2}_{d+2}=\frac{4L^{2}}{(1-u^{2})^{2}}\left[H^{2}u^{2}\left(-(1-H^{2}r^{2})dt^{2}+\frac{dr^{2}}{1-H^{2}r^{2}}+r^{2}d\Omega_{d-1}^{2}\right)+du^{2}\right]\;.\label{eq:statpatchfol}\eeq
which may be transformed into one another via the coordinate change (\ref{eq:coordchglobstat}).

\subsection{Membranes in de Sitter space}

Let us now consider a $(d+1)$-dimensional membrane embedded in $\text{AdS}_{d+2}$ foliated by $\text{dS}_{d+1}$ slices in global coordinates (\ref{eq:globaldSfol}).\footnote{For dS spacetimes, a constant `electric' field is only a solution of the homogeneous gauge field equations when $p=d$. Hence, in this section we specialize to $(d+1)$-dimensional bulk membranes.} In particular, we would like to find membrane solutions preserving the symmetries of a S$^{d-1}$. To do so, we look for embeddings depending only on the polar angle in the line element $d\Omega_{d}^{2}=d\phi^{2}+\sin^{2}\phi d\Omega_{d-1}^{2}$. Parametrizing the worldvolume $\Sigma$ with coordinates $\xi^{i}=(\tau,\phi,\vec{\Omega}_{d-1})$, and choosing the embedding functions to be $X^{\mu}=(\tau,\phi,\vec{\Omega}_{d-1},u(\tau,\phi))$, it is a simple exercise to show the membrane action (\ref{eq:NGactgenp}) becomes
\beq
\begin{split}
S_{d}&
=2^{d+1}T_{d}L^{d+1}\Omega_{d-1}\int d\tau d\theta\frac{u^{d}\cos^{d-1}(H\tau)\sin^{d-1}\phi\sqrt{u'^{2}+\cos^{2}\tau(u^{2}-\dot{u}^{2})}}{(1-u^{2})^{d+1}}\;,
\end{split}
\eeq
where $u'\equiv\frac{du}{d\phi}$ and $\dot{u}\equiv\frac{du}{d\tau}$. From this action one may obtain the membrane equation of motion and solve it for  $u(\tau,\phi)$ subject to appropriate boundary conditions.

We are interested in obtaining membranes expanding at a constant proper acceleration. However, since we have a non-trivial boundary metric, we need to determine the form of such a trajectory, and then impose it as a boundary condition for the bulk membrane. Without lose of generality, we can parametrize the boundary trajectory as $x^{\mu}=\{\tau,\phi(\tau), \Omega_{d-1}\}$. A short calculation reveals that the trajectories we are looking for are of the form
\be\label{trajectdS}
\phi(\tau)=\arccos\left(\kappa\,\text{sech}(H\tau)\right)\,,
\ee
where $\kappa$ is an arbitrary constant. For later convenience, we redefine this constant such that $\kappa=\cos(\alpha)$, in terms of which the norm of the acceleration vector yields
\be\label{accelerationdS}
\sqrt{a_\mu a^\mu}=H \cot(\alpha)\,,\qquad\quad
a^\mu\equiv \frac{d^2x^\mu}{d\tau^2}+\Gamma^{\mu}_{\,\,\,\alpha\beta}\frac{dx^\alpha}{d\tau}\frac{dx^\beta}{d\tau}\,.
\ee
The geometric meaning of $\alpha$ will be clear momentarily, once we move to Euclidean signature.

Imposing (\ref{trajectdS}) as a boundary condition for the bulk membrane leads to the solution
\beq u(\tau,\phi)=\frac{\cosh(H\tau)\cos(\phi)-\sqrt{\cosh^{2}(H\tau)\cos^{2}(\phi)-\cos^{2}(\alpha)}}{\cos(\alpha)}\;,\label{eq:membranesolndSv1}\eeq
where one can check one recovers (\ref{trajectdS}) at $u\to1$. It is worth noting that, similar to the flat space case (\ref{eq:zsolnflatmem}), the embedding (\ref{eq:membranesolndSv1}) for a spherical membrane in dS turns out to be independent of $d$.\footnote{This solution was derived in \cite{Fischler:2014ama} for the case of a string ($d=1$ case), though the angle $\theta$ there was taken to be an azimuthal angle of global dS, rather than a polar angle.}
The solution (\ref{eq:membranesolndSv1}) directly follows from solving the equations of motion for $u(\tau,\phi)$, however, the analysis is cumbersome. In the next subsection we will provide an alternative and far simpler derivation of (\ref{eq:membranesolndSv1}).

Before proceeding further, it is illustrative to translate the membrane solution to static patch coordinates using the transformations (\ref{eq:coordchglobstat}), such that
\beq u(t,r)=\frac{1}{\cos(\alpha)}\left[\sqrt{1-H^{2}r^{2}}\cosh(Ht)-\sqrt{(1-H^{2}r^{2})\cosh^{2}(Ht)-\cos^{2}(\alpha)}\right]\;.\eeq
Alternatively, in terms of the original bulk coordinate $z$ (\ref{eq:coordzr}), we obtain
\beq z(t,r)=\frac{2}{H}\sqrt{\frac{\sqrt{1-H^{2}r^{2}}\cosh(Ht)-\sqrt{1-H^{2}R^{2}}}{\sqrt{1-H^{2}r^{2}}\cosh(Ht)+\sqrt{1-H^{2}R^{2}}}}\;,\qquad\quad R\equiv \frac{\sin(\alpha)}{H}\;.\label{eq:zfuncxt}\eeq
Notice in the flat space limit, $H\to0$, we recover the solution for a spherical membrane in flat space (\ref{eq:zsolnflatmem}) (with $z_m=0$).
Evaluating this solution at $z=0$ we recover
\be\label{staticPtraj}
r(t)=\frac{\text{sech}(Ht)}{H}\sqrt{\sin^{2}(Ht)+H^{2}R^{2}}\;,
\ee
which describes the uniformly accelerated trajectory in the static patch of dS. From the boundary perspective, the membrane contracts, reaches a minimum radius $r=R$ and then expands back at a `constant' rate. However, as seen from the static observer, this rate of expansion is redshifted; this means that for such an observer, the membrane appears to slow down and only reach the cosmological horizon at $r=1/H$ as $t\to\infty$.

\begin{figure}[t!]
\centering
\includegraphics[width=1.8in]{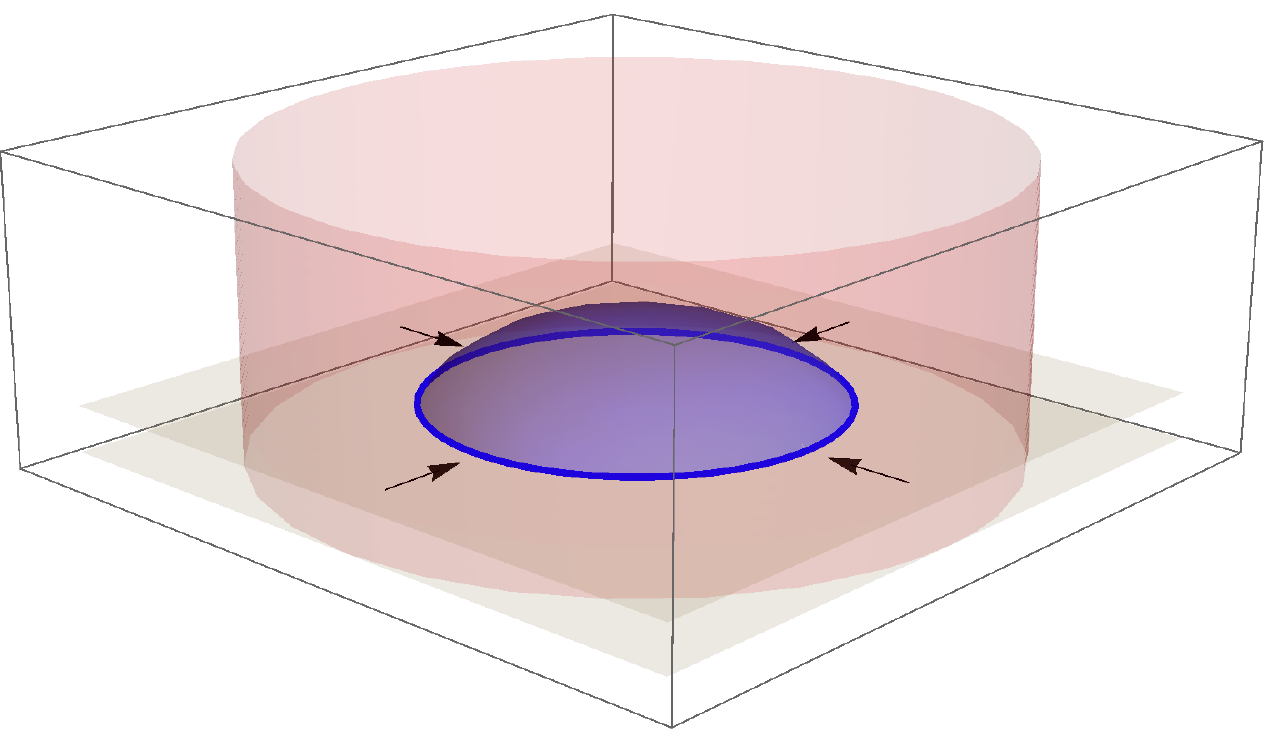}$\qquad$\includegraphics[width=1.8in]{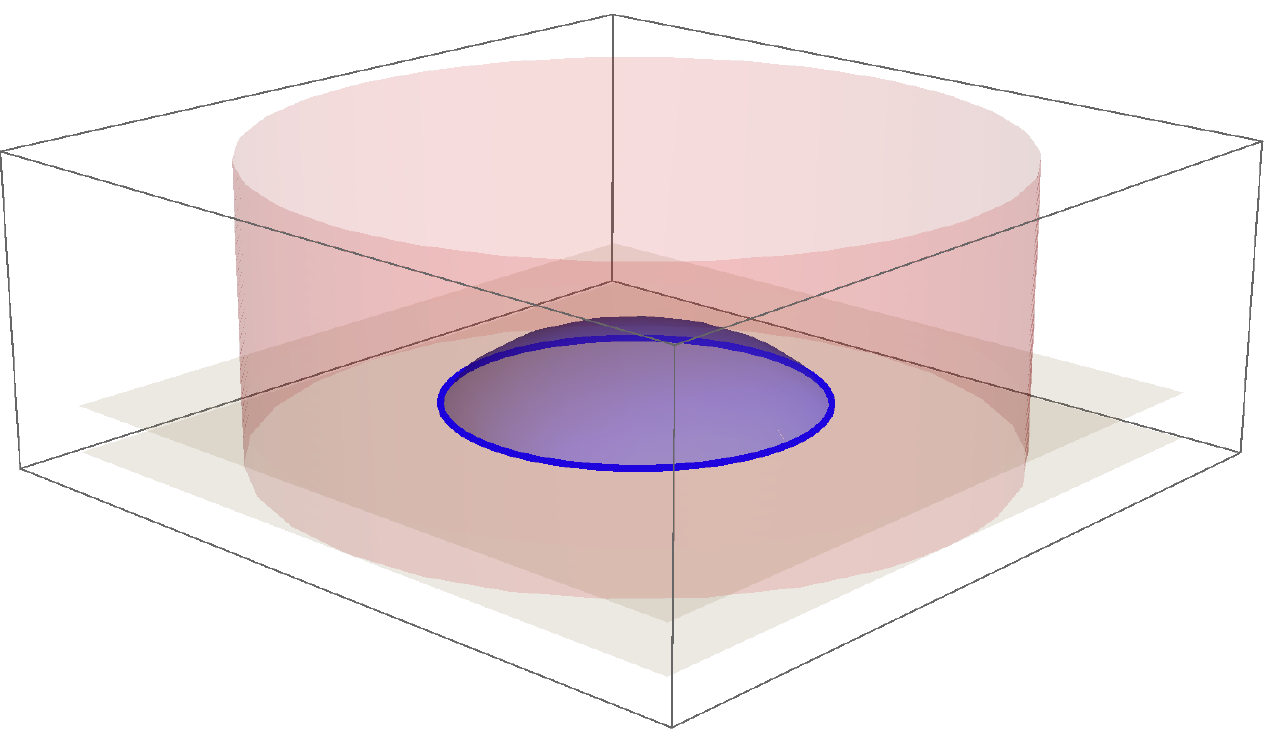}$\qquad$\includegraphics[width=1.8in]{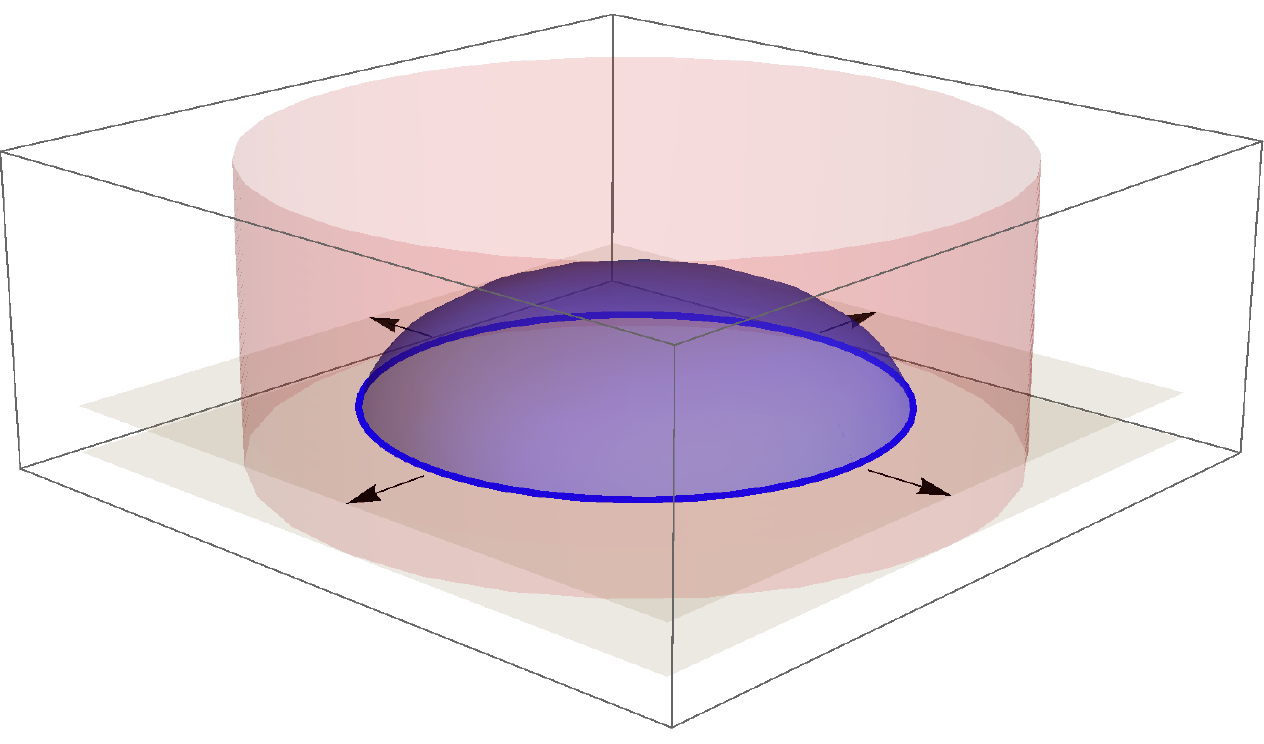}
 \put(-403,9){\tiny $x_1$}
\put(-251,9){\tiny $x_1$}
\put(-99,9){\tiny $x_1$}
 \put(-332,11){\tiny $x_2$}
\put(-180,11){\tiny $x_2$}
\put(-28,11){\tiny $x_2$}
 \put(-327,40){\tiny $rH=1$}
\put(-175,40){\tiny $rH=1$}
\put(-23,40){\tiny $rH=1$}
 \put(-440,43){\tiny $z$}
 \put(-288,43){\tiny $z$}
 \put(-136,43){\tiny $z$}
 \put(-73,20){\tiny $z=z_m$}
\put(-225,20){\tiny $z=z_m$}
\put(-377,20){\tiny $z=z_m$}
 \put(-75,57){\tiny $zH=2$}
\put(-227,57){\tiny $zH=2$}
\put(-379,57){\tiny $zH=2$}
 \put(-68,-5){\tiny $t>0$}
\put(-220,-5){\tiny $t=0$}
\put(-373,-5){\tiny $t<0$}
 \caption{Snapshots of the (Lorentzian) membrane solution (\ref{eq:zfuncxt2}) as it contracts and expands in time. The cosmological and bulk (acceleration) horizons, at $r=1/H$ and $z=2/H$, respectively, are depicted in red. For the sake of the plots, we have only shown the case of a $(2+1)$-dimensional membrane (\emph{i.e.}, with $d=2$), but the solution is valid for arbitrary $d$, with the $d=1$ case reducing to the string worldsheet studied in \cite{Fischler:2014ama}.
\label{figdSLor}}
\end{figure}
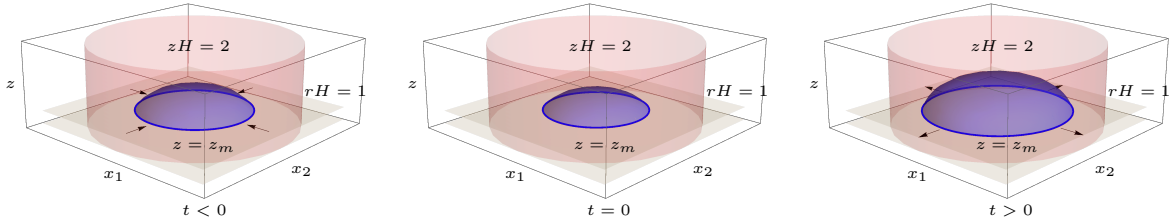

Similar to the flat space case, we now need to truncate this solution at some $z=z_m$. This can be achieved by redefining the constant $R$ according to
\be
R\to\frac{\sqrt{R^{2}(4-H^{2}z_{m}^{2})^{2}+16 z_{m}^{2}}}{(4+H^{2}z_{m}^{2})}\,,
\label{eq:redefR}\ee
so that one recovers (\ref{staticPtraj}) by evaluating the embedding at $z=z_m$. Explicitly, after this substitution, the truncated solution becomes
\be
z(t,r)=\frac{2}{H}\sqrt{\frac{(4+H^2z_m^2)\sqrt{1-H^{2}r^{2}}\cosh(Ht)-(4-H^2z_m^2)\sqrt{1-H^{2}R^{2}}}{(4+H^2z_m^2)\sqrt{1-H^{2}r^{2}}\cosh(Ht)+(4-H^2z_m^2)\sqrt{1-H^{2}R^{2}}}}\;.\label{eq:zfuncxt2}\eeq
Contrary to the flat space case, this solution does not preserve spherical symmetry in the bulk, but rather, describes a membrane that is squashed along the holographic $z$-direction. We can understand this as a result of the non-trivial foliation of AdS. In Figure \ref{figdSLor} we show a few snapshots of the Lorentzian evolution of this solution.
\subsubsection*{Rest energy of membranes}

Let us now compute the energy $\mathcal{E}$ of some membrane solutions,
as we did for a flat boundary.

First, note we can only generalize the static membranes with $x_1=$ constant if we consider the conformally flat patch of dS. A similar solution will not exist in global dS or the static patch, given the lack of invariance under spatial translations. One problem, however, is that it is only in the static patch that we have a time translation symmetry, so that energy can be properly defined. Now, focusing on the static patch, one may be tempted to look for static solutions that reach the boundary at some $r=$ constant. This is indeed possible for $d=1$, as shown in \cite{Fischler:2014ama}; however, such a solution does not generalize for higher $d$. Solving perturbatively close to the boundary, one quickly learns that the embedding function becomes imaginary, meaning that one needs to look for a more general ansatz (\emph{i.e.}, one that allows for time-dependence in the bulk while keeping the radius fixed at the boundary). We will not do this here. Instead, we will only consider the membrane solutions we have already derived, dual to a boundary membrane contracting and expanding at a constant proper acceleration.

We thus consider the AdS$_{d+2}$ metric foliated with dS$_{d+1}$ slices in the static patch. Parametrizing the worldvolume with coordinates $\xi^{i}=(t,z,\vec{\Omega}_{d-1})$ and choosing the embedding as $X^{\mu}=(t,r(t,z),\vec{\Omega}_{d-1})$, the membrane action becomes\footnote{Note one can easily invert (\ref{eq:zfuncxt2}) to obtain $r(t,z)$.}
\beq S_{d}=T_{d}\Omega_{d-1}\int dzdt\left(\frac{Lfr}{z}\right)^{d-1}\sqrt{\dot{X}^{2}X'^{2}-(\dot{X}\cdot X')^{2}}\;.\eeq
Likewise, in this parametrization, the energy density $\Pi_{t}$ reads
\beq \Pi_{t}=T_{d}L^{d+1}\Omega_{d-1}\frac{f^{d}r^{d-1}}{z^{d+1}}\left(\frac{h(r)+f^{2}r'^{2}}{\sqrt{h(r)+f^{2}r'^{2}-\frac{\dot{r}^{2}}{h(r)}}}\right)\;,\label{eq:EndendSmem}\eeq
with $h(r)\equiv 1-H^{2}r^{2}$. The energy $\mathcal{E}$ then follows from integrating $\Pi_{t}$ with respect to $z$ from the cutoff $z_{m}$ to the maximum value of $z$ (which occurs at $r=0$), leading to
\beq \mathcal{E}=T_{d}L^{d+1}\Omega_{d-1}\frac{\sqrt{1-H^{2}R^{2}}(4-H^{2}z_{m}^{2})^{d+1}}{d\sqrt{16z_{m}^{2}+R^{2}(4-H^{2}z_{m}^{2})^{2}}}\left(\frac{\sqrt{\sinh^{2}(Ht)+H^{2}R^{2}}}{4Hz_{m}\cosh(Ht)}\right)^{d}\;.\label{eq:enmemdS}\eeq
Finally, the rest energy of the membrane $\mathcal{E}_{0}\equiv\mathcal{E}(t=0)$ is
\beq
\mathcal{E}_{0}=\frac{T_{d}\Omega_{d-1}L^{d+1}}{d}\frac{\sqrt{1-H^{2}R^{2}}f(z_{m})^{d+1}}{\sqrt{z_{m}^{2}+R^{2}f(z_{m})^{2}}}\left(\frac{R}{z_{m}}\right)^{d}\;,
\label{eq:restenmemdS}\eeq
with $f(z_{m})=1-\frac{H^{2}z_{m}^{2}}{4}$. Note that in the $H\to0$ limit, we recover the energy of a spherical membrane in flat space (\ref{eq:E0flatmemexact}). Further, as explained previously, we expect any $z_m\neq0$ to lead to a boundary membrane with finite width, departing from the infinitely thin shell regime.

\subsection{On-shell Euclidean action and nucleation rate}

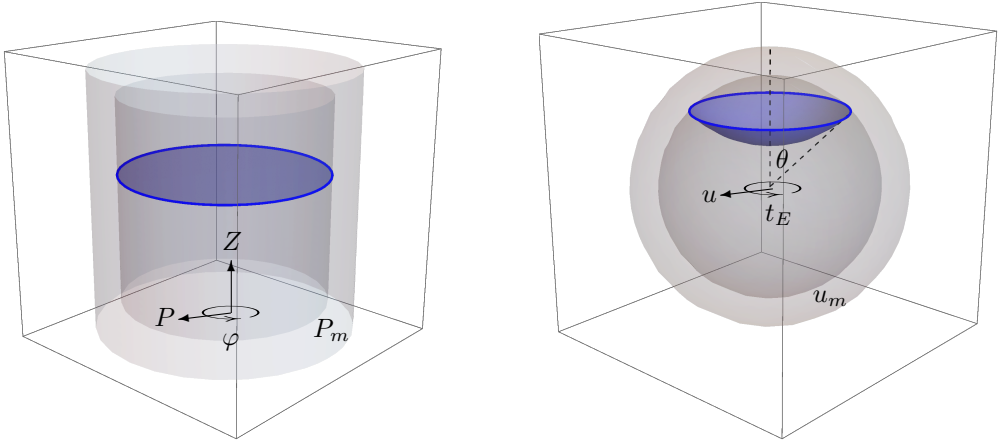
\begin{figure}[t!]
\centering
\includegraphics[width=2.4in]{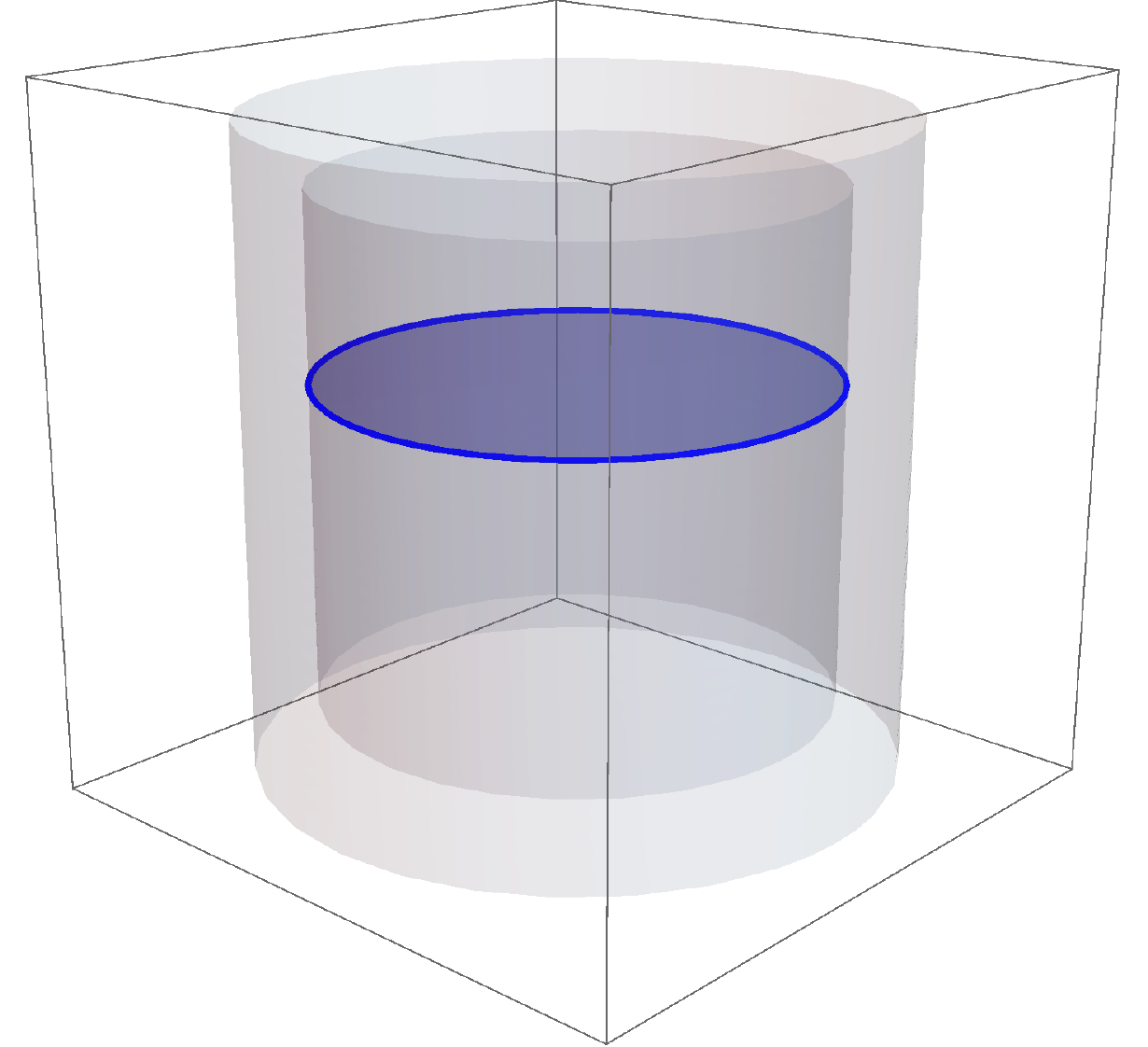}$\qquad\quad$\includegraphics[width=2.4in]{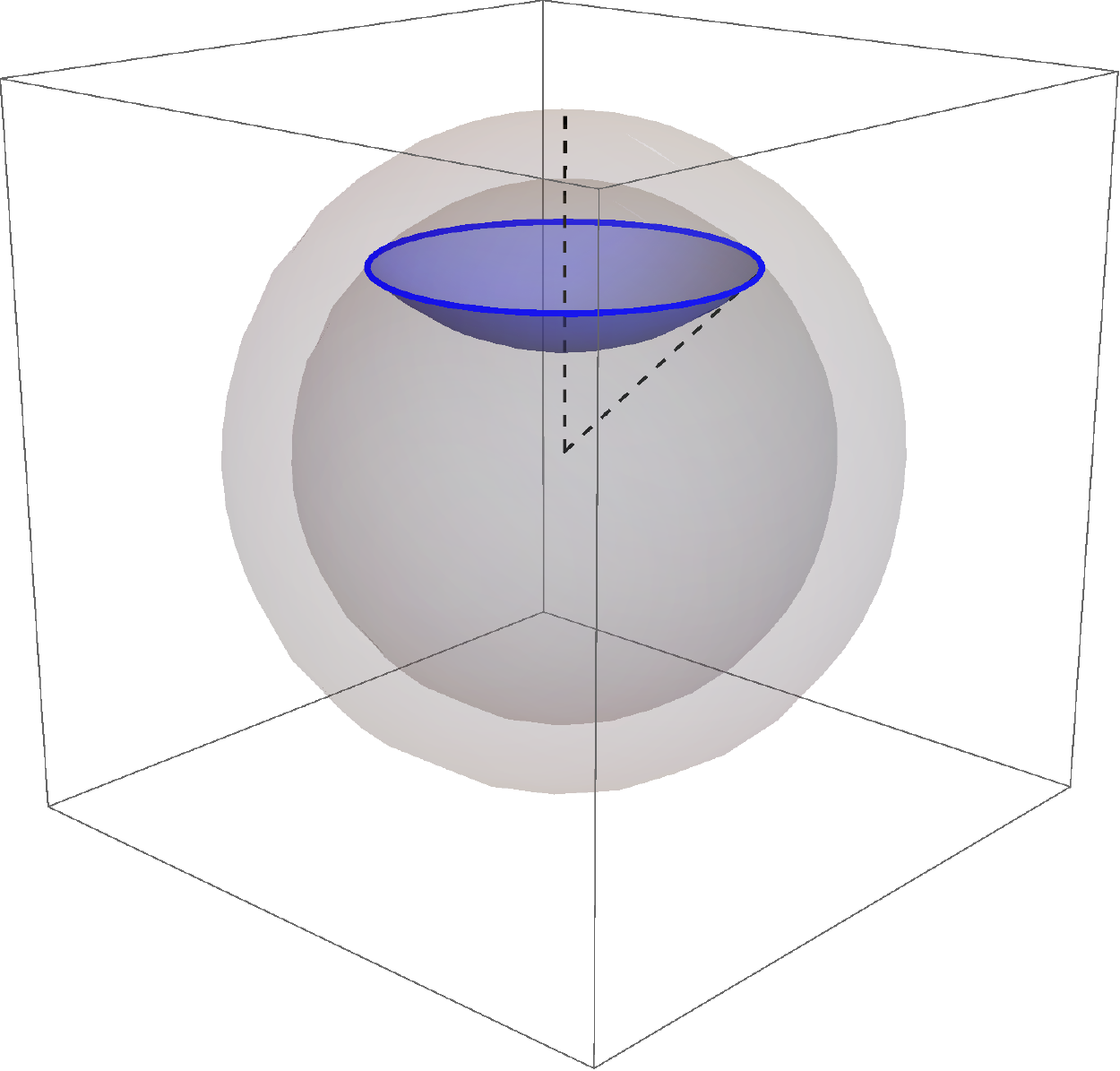}
\put(-290,48){\vector(0,1){20}}
\put(-310,48){\rotatebox{97}{\vector(0,1){20}}}
\put(-306,51){\scalebox{3}[.7]{\rotatebox{200}{$ \circlearrowleft$}}}
\put(-293.5,72){\small $Z$}
\put(-319,43){\small $P$}
\put(-293.5,36){\small $\varphi$}
\put(-259,39){\small $P_m$}
\put(-105,95){\rotatebox{97}{\vector(0,1){20}}}
\put(-101,98){\scalebox{3}[.7]{\rotatebox{200}{$ \circlearrowleft$}}}
\put(-88,83){\small $t_E$}
\put(-112,90){\small $u$}
\put(-84,103){\small $\theta$}
\put(-70,52){\small $u_m$}
 \caption{Left: Membrane solution (\ref{eq:solcylind}) in the Euclidean cylinder with bulk cutoff $P_{m}$. Right: Membrane solution (\ref{eq:Eucsolnmem}) in the Euclidean ball with bulk cutoff $u_{m}$. The two solutions map to each other under the bulk transformations (\ref{eq:PZphicoord}). Furthermore, to properly account for the finite cutoff, one needs to redefine $\alpha$ according to (\ref{eq:anglereplace}). \label{figdSM}}
\end{figure}

Let us now turn to the on-shell action for spherical membranes nucleating in $\text{dS}_{d+1}$. Our starting point is the Euclideanized action (\ref{eq:totalactEuc}) including the gauge field contribution $S_{E}^{A}$. To evaluate the Euclidean action, it is most convenient to work in coordinates where Euclidean AdS is represented by a solid cylinder \cite{Krtous:2014pva,Fischler:2014ama}
\beq ds^{2}_{d+2}=L^{2}\left[\frac{dP^{2}}{1+P^{2}}+(1+P^{2})dZ^{2}+P^{2}d\Omega_{d}^{2}\right]\;.\label{eq:AdSascyl}\eeq
with $\vec{\Omega}_d=\{\varphi,\vec{\Omega}_{d-1}\}$. The coordinates $(P,Z,\varphi)$ are related to Euclidean $\text{AdS}_{d+2}$ foliated by static patch coordinates via (see Appendix \ref{app:dSfoliationsAdS})
\beq
\begin{split}
&P=\frac{2u}{1-u^{2}}\sqrt{1-\cos^{2}t_{E}\cos^{2}\theta}\;,\\
&Z=\text{arccosh}\left(\frac{1+u^{2}}{\sqrt{(1+u^{2})^{2}-4u^{2}\cos^{2}t_{E}\cos^{2}\theta}}\right)\;,\\
&\varphi=\text{arccos}\left(\frac{\cos\theta\sin t_{E}}{\sqrt{1-\cos^{2}t_{E}\cos^{2}\theta}}\right)\;,
\end{split}
\label{eq:PZphicoord}\eeq
where we have Wick rotated the time $t$ of a static observer $t\to -i t_{E}/H$, and where the angular variable $\theta$ is related to the radial coordinate $r$ via $Hr=\sin\theta$, such that the cosmological horizon is located at $\theta=\pi/2$.\footnote{The remaining angular coordinates, $\vec{\Omega}_{d-1}$, remain untouched.} In these coordinates the conformal boundary is located at $P\to\infty$, however, it is typically cut off by a surface at $P=P_m$. See Figure \ref{figdSM} (left).

We choose to parametrize the worldvolume $\Sigma_d$ with coordinates $\xi^{i}=(P,\vec{\Omega})$, and pick embedding functions $X^{\mu}=(P,\vec{\Omega}_{d},Z(P))$. Consequently, the Euclidean membrane action is
\beq S^{E}_{d}=T_{d}L^{d+1}\Omega_{d}\int_{\Sigma_{d}} dPP^{d}\sqrt{\frac{(1+P^{2})^{2}Z'^{2}+1}{1+P^{2}}}\;.\label{eq:memactdSEuc}\eeq
It is straightforward to show that $Z(P)=$ constant is a solution to the equations of motion for $Z$. In particular, it is natural to fix
\beq Z(P)=\frac{1}{\sin\alpha}\;,\label{eq:solcylind}\eeq
such that $\alpha$ coincides the instanton's angle on the Euclidean ball with respect to the axis of symmetry.\footnote{We will shortly truncate this solution at the surface $P=P_m$, which will induce a shift in $\alpha$.} To see this, we transform the solution (\ref{eq:solcylind}) to static patch coordinates  (\ref{eq:dSstatEucapp}),
\beq u(\t_{E},\theta)=\frac{\cos t_{E}\cos\theta-\sqrt{\cos^{2} t_{E}\cos^{2}\theta-\cos^{2}\alpha}}{\cos\alpha}\;.\label{eq:Eucsolnmem}\eeq
We recognize this as the analytic continuation of the Lorentzian solution (\ref{eq:membranesolndSv1}). Further, we note the parameter $\alpha$ has a very clear geometric meaning in this coordinate system, as the polar angle subtended by the membrane at the conformal boundary, $u\to1$. See Figure \ref{figdSM} (right) for an illustration.

As before, we need to truncate these solution at the cutoff surface $P=P_{m}$ to avoid a diverging mass of the membrane. We may relate $P_{m}$ to the original cutoff $u_{m}$ (or, equivalently, $z_m$) by inverting (\ref{eq:Eucsolnmem}) to express the angular variables in terms of $u$ and $\alpha$, and substitute the result into the definition of $P$ (\ref{eq:PZphicoord}), yielding
\beq P_{m}=\frac{2u_{m}}{(1-u_{m}^{2})}\sqrt{1-\frac{(1+u_{m}^{2})^{2}}{4u_{m}^{2}}\cos^{2}(\alpha)}\;,\label{eq:Pmfirstpass}\eeq
where the cutoffs $u_{m}$ and $z_{m}$ are related via
\beq \frac{(1+u_{m}^{2})}{2u_{m}}=\left(\frac{4+H^{2}z_{m}^{2}}{4-H^{2}z_{m}^{2}}\right)\;.\label{eq:alphaalpham}\eeq
We can think of $\alpha$ as the polar angle on the $(d+1)$-dimensional (Euclidean) de Sitter sphere of radius $H^{-1}$, such that the radius of the membrane instanton is $R=H^{-1}\sin(\alpha)$. In order to consistently truncate the solution at the cutoff surface, we thus need to redefine this radius according to the replacement rule (\ref{eq:redefR}). In terms of $\alpha$, this amounts to replacing
\beq \cos(\alpha)\to\left(\frac{4-H^{2}z_{m}^{2}}{4+H^{2}z_{m}^{2}}\right)\cos(\alpha)\;,\label{eq:anglereplace}\eeq
which, via (\ref{eq:coordzr}) and (\ref{eq:Pmfirstpass}), implies
\beq P_{m}=\frac{2u_{m}}{(1-u_{m}^{2})}\sin(\alpha)=\left(\frac{1}{Hz_{m}}-\frac{Hz_{m}}{4}\right)\sin(\alpha)\;.\label{eq:Pm}\eeq

We are now ready to compute the on-shell action. Using $Z'=0$, the Euclidean membrane action (\ref{eq:memactdSEuc}) may be computed exactly in any number of dimensions $d$,
\beq S_{d}^{E}=T_{d}L^{d+1}\Omega_{d}\int_{0}^{P_{m}}dP\frac{P^{d}}{\sqrt{1+P^{2}}}=T_{d}L^{d+1}\Omega_{d}\frac{P^{d+1}_{m}}{(d+1)}{}_{2}F_{1}\left(\frac{1}{2}\,,\,\frac{d+1}{2}\,,\,\frac{d+3}{2}\,,\,-P_{m}^{2}\right)\;.\eeq
Using (\ref{eq:Pm}), we may rewrite the result in terms of $z_{m}$ and $R$, so that
\beq S_{d}^{E}=\frac{T_{d}L^{d+1}\Omega_{d}}{(d+1)}\left(\frac{R}{z_{m}}\right)^{d+1}f(z_{m})^{d+1}{}_{2}F_{1}\left(\frac{1}{2}\,,\,\frac{d+1}{2}\,,\,\frac{d+3}{2}\,,\,-\frac{R^{2}}{z_{m}^{2}}f(z_{m})^{2}\right)\;.\eeq
Furthermore, as in the flat space case, the on-shell Euclidean action for the external gauge field $A$ is given by
\beq S^{E}_{A}=\int_{\partial\Sigma_{d}}A=-E\int_{\mathcal{V}_{d+1}}\tilde{\epsilon}=-E\mathcal{V}_{d+1}\;,\eeq
with $\mathcal{V}_{d+1}$ being the Euclidean spacetime volume enclosed by the boundary of the membrane with radius $R=H^{-1}\sin(\alpha)$,
\beq \!\mathcal{V}_{d+1}=\frac{\Omega_{d}}{H^{d+1}}\int_{0}^{\alpha}\!\sin^{d}\theta \,d\theta=\frac{\Omega_{d}}{H^{d+1}}\left[\frac{\sqrt{\pi } \Gamma [\frac{d+1}{2}]}{2 \Gamma [\frac{d}{2}+1]}-\cos (\alpha ) \!\, _2F_1\left(\frac{1}{2},\frac{1-d}{2},\frac{3}{2},\cos ^2(\alpha )\right)\right].\eeq
Combining these two results, we obtain the total Euclidean action $S_{E}$ yields
\beq
\begin{split}
 S_{E}&=S_{d}^{E}+S^{E}_{A}\\
 &=\frac{T_{d}L^{d+1}\Omega_{d}}{(d+1)}\left(\frac{R}{z_{m}}\right)^{d+1}f(z_{m})^{d+1}{}_{2}F_{1}\left(\frac{1}{2}\,,\,\frac{d+1}{2}\,,\,\frac{d+3}{2}\,,\,-\frac{R^{2}}{z_{m}^{2}}f(z_{m})^{2}\right)-E\mathcal{V}_{d+1}\;.
\end{split}
\label{eq:totalEucact}\eeq
For example,
\beq
\begin{split}
&S_{E}^{(d=1)}=\sqrt{\lambda}\left(\sqrt{1+\left(\frac{1}{Hz_{m}}-\frac{Hz_{m}}{4}\right)^{2}\sin^{2}(\alpha)}-1\right)-\frac{2\pi E}{H^{2}}\left(1-\cos(\alpha)\right)\;,\\
&S_{E}^{(d=2)}=2\pi L^{2}T_{2}\left\{\frac{\sin(\alpha)}{Hz_{m}}f(z_{m})\sqrt{1+\left(\frac{\sin(\alpha)}{Hz_{m}}\right)^{2}f(z_{m})^{2}}-\text{arcsinh}\left[\frac{\sin(\alpha)}{Hz_{m}}f(z_{m})\right]\right\}\\
&\qquad\quad\;\;\;-\frac{2\pi E}{H^{3}}\left[\alpha-\cos(\alpha)\sin(\alpha)\right]\;,\\
&S_{E}^{(d=3)}=\frac{2\pi^{2}}{3}L^{4}T_{3}\left\{2+\sqrt{1+\left(\frac{\sin(\alpha)}{Hz_{m}}\right)^{2}f(z_{m})^{2}}\left[\left(\frac{\sin(\alpha)}{Hz_{m}}\right)^{2}f(z_{m})^{2}-2\right]\right\}\\
&\qquad\quad\;\;\;-\frac{8\pi^{2}E}{3H^{4}}\left(2+\cos(\alpha)\right)\sin^{4}\left(\frac{\alpha}{4}\right)\;,
\end{split}
\label{eq:d123dSact}\eeq
where in the first line we used $T_{1}=\frac{\sqrt{\lambda}}{2\pi L^{2}}$, recovering the string result derived in \cite{Fischler:2014ama}. Note that setting $\sin\alpha=HR$ and taking the limit $H\to0$, we exactly recover the on-shell action for spherical membranes in flat space (\ref{eq:flatmemsd123act}).

We can fix the radius $R$ (or, equivalently, $\alpha$) by extremizing the on-shell action (\ref{eq:totalEucact}) with respect to $\alpha$. A quick calculation yields
\beq \sin^{2}(\alpha)=\frac{(4-H^{2}z_{m}^{2})^{2(d+1)}(HL)^{2(d+1)}T_{d}^{2}-(4Hz_{m})^{2(d+1)}E^{2}}{(4-H^{2}z_{m}^{2})^{2}[(HL)^{2(d+1)}T_{d}^{2}+(4Hz_{m})^{2d}E^{2}]}\;.\label{eq:angleapham}\eeq
As a consistency check, notice one may recover (\ref{eq:Rrad}) in the $H\to0$ limit. Further, the associated critical value for the field magnitude $E_{c}$, which occurs when $R\to0$, is given by
\beq E_{c}=T_{d}\left(\frac{L}{z_{m}}\right)^{d+1}\left(1-\frac{H^{2}z_{m}^{2}}{4}\right)^{d+1}\;,\label{eq:EcritdS}\eeq
which differs from the flat space result (\ref{eq:Ecritflat}) only by an overall factor of $f(z_{m})^{d+1}$. In terms of $E_{c}$, the radius $R$ is
\beq R=H^{-1}\sin(\alpha)=\frac{4^{d}z_{m}}{f(z_{m})}\sqrt{\frac{(E_{c}^{2}-E^{2})}{[(Hz_{m})^{2}f(z_{m})^{-2(d+1)}E_{c}^{2}+4^{2d}E^{2}]}}\;,\label{eq:raddSEc}\eeq
which indeed vanishes in the limit $E\to E_{c}$. Since the total action $S_{E}$ vanishes when $R$ (or $\alpha$) goes to zero, the action will likewise  vanish in the limit $E\to E_{c}$, indicating an unstable vacuum.

It is worth emphasizing that in the $E\to0$ limit, \emph{i.e.}, when the external field $A$ is turned off, the on-shell action $S_{E}$ remains finite, unlike the flat space case, where it diverges. This tells us that membrane nucleation occurs in de Sitter space when there is no external gauge field present, mediated solely by the background gravitational field. This scenario corresponds to the spontaneous nucleation of membranes or topological defects due to the accelerated expansion present in the de Sitter background, analogous to \cite{Basu:1991ig,Basu:1992ue}. Thus, the on-shell action (\ref{eq:totalEucact}) with $E=0$ provides the dominant contribution to the nucleation of defects during inflation, at strong coupling.

Finally, with some effort, in the weak field limit $E \ll E_c$ and $z_m \ll R$ (or, equivalently, $\mathcal{E}_{0}\gg T_{d}L^{d+1}H$), one exactly recovers the weakly coupled result (\ref{eq:sphericalmemactapp}) first uncovered in \cite{Garriga:1993fh}. Explicitly,
\beq S_{E}=\frac{\sqrt{\pi}\Gamma\left(\frac{d}{2}\right)}{\Gamma\left(\frac{d+1}{2}\right)}\frac{\mathcal{E}_{0}R}{\sqrt{1-H^{2}R^{2}}}-qE\Omega_{d}H^{-(d+1)}\int_{0}^{\alpha}d\theta\sin^{d}(\theta)\;,\eeq
where $\sin(\alpha)=HR$ and $R=(d\mathcal{E}_{0}/qE\Omega_{d-1})^{1/d}$. As in the flat space case, this weak coupling result receives corrections represented by higher powers in $E/E_0$ and $T_{d}L^{d+1}H/\mathcal{E}_{0}$, which may be regarded as non-perturbative corrections due to strong
coupling and strong electric fields, and away from the infinitely thin shell regime.

%%%%%%%%%%%%%%%%%%%%%%%%%%%%%%%%%%%%%%%%%%%%%%%%%%%%%%%
\section{Conclusion} \label{sec:conclusion}

In this article we computed the nucleation rates of spherical membranes in flat and de Sitter backgrounds at strong coupling using AdS/CFT. This was accomplished using instanton techniques by computing the on-shell actions describing the worldvolume of membranes ending on a cutoff surface in the $\text{AdS}_{d+2}$ bulk. In the case of nucleation rates in flat space, we considered empty $\text{AdS}_{d+2}$ with a flat conformal boundary, and coupled a bulk ($p+1$)-dimensional membrane ($p\leq d$) to an antisymmetric tensor field with a field strength tensor of constant field magnitude, a higher dimensional analog of the holographic Schwinger effect.
Importantly, the cutoff surface was taken to be at a finite distance of the conformal boundary of AdS so, via the standard UV/IR connection of AdS/CFT, the dual $p$-dimensional membranes being nucleated at the boundary are thus finite width membranes. Similarly, by considering $\text{dS}_{d+1}$ foliations of $\text{AdS}_{d+2}$, such that the conformal boundary of AdS was de Sitter space, we analytically computed the nucleation rate of $d$-dimensional membranes in $\text{dS}_{d+1}$.\footnote{Crucially, in dS space, a constant electric field $E$ can be supported without charges only for $p=d$.} Via holography, our findings  are inherently at strong coupling. In either set-up, we found exact agreement with prior nucleation rate computations in the weak coupling limit. Beyond the weak coupling limit, we found the nucleation rates to be corrected non-perturbatively, in which higher powers in the field strength and the coupling constant appear, analogous to the holographic Schwinger effect at strong coupling.

There are a number of future research directions worth pursuing.
Firstly, as membranes nucleate, there will be a finite probability for these membranes to collide. Such membrane collisions provide a toy model for bubble collisions arising during vacuum decay \cite{Guth:1982pn,Garriga:2006hw}, for which there exist exact solutions, e.g., \cite{Freivogel:2007fx,Chang:2007eq} and may have observational signatures (cf. \cite{Aguirre:2007an,Aguirre:2007wm,Chang:2008gj}). It would be straightforward to extend our analysis to the case of spherical membrane collisions using the solutions found in \cite{Vegh:2021sva}. Second, it would be interesting to study the nucleation of strings and membranes in other spacetimes, e.g., AdS black hole backgrounds, by considering a different foliation of bulk $\text{AdS}_{d+2}$, along the lines of \cite{Marolf:2013ioa,Sato:2013hyw}.

Thirdly, in this article we only focused on the exponential contribution to the nucleation rate, $\Gamma=A e^{-S_{E}}$, ignoring the details of the prefactor $A$. The prefactor encodes bulk quantum fluctuations about the instantons, and follows from an evaluation of one-loop functional determinants. It would be interesting to study the influence of the quantum fluctuations about the instanton solutions in the holographic model uncovered here, for which the analysis given in \cite{Garriga:1993fh} would be of use. On general grounds, we expect the contributions of these quantum fluctuations to be relevant if we depart from the strict infinite-$N_c$ limit.

Lastly, the creation of membranes considered here may be seen as a toy model for vacuum decay. This follows due to the fact we treated both the antisymmetric tensor field and gravitational field as external, neglecting the effects of backreaction. It would therefore be very interesting to consider full fledged vacuum decay, e.g., the Coleman-De Luccia mechanism \cite{Coleman:1980aw}, in the context of AdS/CFT, where gravity is allowed to be dynamical. There have been a plethora of previous studies on understanding false vacuum decay via AdS/CFT, e.g., \cite{Alberghi:1999kd,Balasubramanian:2002am,Banks:2002nm,Ross:2004cb,Balasubramanian:2005bg,deHaro:2006ymc,Papadimitriou:2007sj,He:2007ji,Barbon:2010gn,Maldacena:2010un,Barbon:2011ta,Ghosh:2021lua,Antonelli:2018qwz}. One way of introducing dynamical gravity is through braneworld holography,\footnote{The study of the Coleman-De Luccia mechanism on a flat Randall-Sundrum braneworld without a holographic description was considered in \cite{Cuspinera:2019jwt}.} where one considers a bulk AdS spacetime with a brane inside with its own induced dynamical gravity (for a recent review see \cite{Chen:2020uac}). A similar set-up was analyzed in \cite{Maldacena:2010un}, where a Minkowski false vacuum decays into AdS, where the decay geometry contains only a portion of AdS, a Coleman-De Luccia AdS bubble. Then, via AdS/CFT, this portion of AdS is dual to a well defined CFT with a cutoff such that the CFT lives on a domain wall, and is spiritually similar to decays due to end of world branes \cite{Witten:1981gj}. Alternatively, one could consider a set-up in which dynamical gravity exists on the conformal boundary of AdS. Such a scenario was developed in \cite{Compere:2008us,Ecker:2021cvz}, where boundary counterterms in the bulk gravity action lead to a class of boundary conditions which render the boundary metric dynamical. The  boundary gravitational dynamics is then induced by the holographic CFT. It would be interesting to push this further to study false vacuum decay \'a la Coleman-De Luccia. Further, a setting with dynamical gravity would allow one to test the quantum nature of dS spacetime, \emph{e.g.}, the imprint of the (possibly) finite dimensional Hilbert space of dS on nucleation rates \cite{Banks:2000fe,Bousso:2000nf,Banks:2001yp,Witten:2001kn,Banks:2003cg,Parikh:2004ux,Susskind:2021dfc,Anninos:2022ujl,Morvan:2022ybp}.

\noindent \noindent\section*{Acknowledgments}
We are grateful to Ruth Gregory, Tanmay Vachaspati, and George Zahariade for useful correspondence. MA is supported by the Mexican National Council of Science and Technology (CONACYT), and by EPSRC.  WF is supported by the National Science Foundation
under Grant Number PHY-1914679. JFP is supported by the `Atracci\'on de Talento' program (2020-T1/TIC-20495, Comunidad de Madrid) and by the Spanish Research Agency (Agencia Estatal de Investigaci\'on) through the Grant IFT Centro de Excelencia Severo Ochoa No. CEX2020-001007-S, funded by MCIN/AEI/10.13039/501100011033. AS is supported by the Simons Foundation via \emph{It from Qubit: Simons Collaboration on quantum fields, gravity, and information}, EPSRC and a MAPS ECR travel grant. AS would also like to thank IFT UAM/CSIC for hospitality while this work was being completed.

%%%%%%%%%%%%%%%%%%%%%%%%%%%%%%%
\appendix

\section{Nucleation at weak coupling} \label{app:nucweak}

Here we review the computation of nucleation rates of membranes in de Sitter space at weak coupling \cite{Garriga:1993fh} without the employ of holography. These results act as a useful consistency check with the strong coupling calculations performed in the main text.

\subsection{Membranes in flat space}

It is illustrative to first consider the creation of spherical membranes in $(d+1)$-dimensional flat space. Analogous to Schwinger pair production, membranes may be created in $(d+1)$-dimensional flat or de Sitter space via an external antisymmetric tensor field $A=A_{[\mu...\rho]}dx^{\mu}\wedge...\wedge dx^{\rho}$ of rank $d$ coupled to a $d$-dimensional worldvolume $\Sigma$ of a charged membrane \cite{Brown:1988kg}. The total Lorentzian action $S$ of the membrane is given by the sum of a generalization of the Nambu-Goto action for a string action coupled to the external tensor field $A$,
\beq S=\mathcal{M}\int_{\Sigma}d^{d}\xi\sqrt{\text{det}\gamma_{ij}}+q\int_{\Sigma}A\;.\label{eq:lorentactmemflat}\eeq
Here $\xi^{i}$ for $i=0,...,d-1$ denote a collection of coordinates parametrizing the worldvolume, $\mathcal{M}$ is a constant representing the tension of the membrane, and $q$ is the membrane charge. The field strength associated with $A$, denoted by $F=dA$, has only a single independent component, such that $F=-E\tilde{\epsilon}$, where $E$ is the analog of the constant electric field in Schwinger pair production, and $\tilde{\epsilon}$ is the spacetime volume form.  For $d=1$, the above action simply characterizes  a spinless particle of mass $\mathcal{M}$ and charge $q$ interacting with a Maxwell field $A$, with $\Sigma$ being the particle's worldline.

One can relate the parameter $\mathcal{M}$ to the (rest) energy $\mathcal{E}_{0}$ of the membrane. To see this, we embed the worldvolume into Minkowski space with embedding coordinates $X^{\mu}=(t,r(t),\vec{\Omega}_{d-1})$, where $\vec{\Omega}_{d-1}$ represents the collection of angular coordinates, and parametrize the worldvolume with coordinates $\xi^{i}=(t,\vec{\Omega}_{d-1})$. The Lorentzian membrane action then is
\beq S=\mathcal{M}\Omega_{d-1}\int_{\Sigma}dt r^{d-1}\sqrt{-g_{\mu\nu}\dot{X}^{\mu}\dot{X}^{\nu}}\;,\eeq
where $\dot{X}^{\mu}\equiv \frac{\partial X^{\mu}}{\partial t}$. The momentum  $\Pi_{\mu}$ then is
\beq \Pi_{\mu}=\frac{\partial\mathcal{L}_{d}}{\partial \dot{X}^{\mu}}=-\mathcal{M}r^{d-1}\Omega_{d-1}\frac{\dot{X}_{\mu}}{\sqrt{-g_{\mu\nu}\dot{X}^{\mu}\dot{X}^{\nu}}}\;,\eeq
for which the energy $\mathcal{E}$
\beq \mathcal{E}\equiv\Pi_{t}=\mathcal{M}\Omega_{d-1}\frac{r^{d-1}}{\sqrt{1-\dot{r}^{2}}}\;.\eeq
The Lorentzian solution for a spherical membrane is given by a worldvolume of radius $R$
\beq r(t)=\sqrt{R^{2}+t^{2}}\;,\label{eq:solflatsol}\eeq
for any $d$. Substituting this into $\Pi_{t}$ yields the energy
\beq \mathcal{E}=\frac{\mathcal{M}\Omega_{d-1}}{R}(R^{2}+t^{2})^{d/2}\;.\eeq
The `rest' energy $\mathcal{E}_{0}=\mathcal{E}(t=0)$ is then
\beq \mathcal{E}_{0}=\mathcal{M}\Omega_{d-1}R^{d-1}\;.\label{eq:Enmemflatapp}\eeq

\subsubsection*{Instantons}

The nucleation rate $\Gamma$ of these spherical membranes are described by instantons, solutions to the Euclidean action, such that $\Gamma\sim e^{-S_{E}}$, where $S_{E}$ is the on-shell Euclidean action.  The total Euclidean action $S_{E}$ of the membrane follows from Wick rotating $t\to -it_{E}$ the Lorentzian action\footnote{One also uses that the field strength $F$ is unchanged under Wick rotation, which requires Wick rotating the \emph{spatial} components of the gauge field $A$.} (\ref{eq:lorentactmemflat}),
\beq S_{E}=\mathcal{M}\int_{\Sigma}d^{d}\xi\sqrt{\text{det}\gamma_{ij}}+q\int_{\Sigma}A\;.\label{eq:IEmemgarri}\eeq
Since the membrane is closed, by Stokes' theorem the Euclidean action (\ref{eq:IEmemgarri}) becomes
\beq S_{E}=\mathcal{M}\int_{\Sigma}d^{d}\xi\sqrt{\text{det}\gamma}-qE\int_{\mathcal{V}}\tilde{\epsilon}\;.\label{eq:IEmemgarriv2}\eeq

Directly from the action $S_{E}$ (\ref{eq:IEmemgarriv2}), the Euclidean equations of motion of the action (\ref{eq:IEmemgarriv2}) were computed in \cite{Garriga:1991ts,Guven:1993ew} and are characterized by the extrinsic curvature $K_{ij}$ of the Euclidean worldvolume,
\beq \gamma^{ij}K_{ij}=-\frac{qE}{\mathcal{M}}\;.\eeq
Spherical worldvolumes of radius $R$ with solution $r(t_{E})=\sqrt{R^{2}-t_{E}^{2}}$, where
\beq R=\frac{d\mathcal{M}}{qE}=\left(\frac{d\mathcal{E}_{0}}{qE\Omega_{d-1}}\right)^{1/d}\;,\label{eq:flatrad}\eeq
 follows from extremizing the Euclidean action. Upon Wick rotating back to Lorentzian signature, one recovers the solution (\ref{eq:solflatsol}). Consequently, the on-shell Euclidean action for spherical membranes in $(d+1)$-dimensional flat space follows from substituting $r(t_{E})$ into (\ref{eq:IEmemgarriv2}) where $t_{E}\in[-R,R]$ for the integration range, leading to:
\beq
\begin{split}
 S_{E}&=\mathcal{M}S_{d}(R)-qE\mathcal{V}_{d}=\frac{\Omega_{d}\mathcal{M}R^{d}}{(d+1)}=\frac{\Omega_{d}}{(d+1)}\left[\frac{d}{qE}\left(\frac{\mathcal{E}_{0}}{\Omega_{d-1}}\right)^{(d+1)}\right]^{1/d}\;.
\end{split}
\label{eq:Euconshellflatapp}\eeq
Here $S_{d}$ is the surface area of $\Sigma_{d}$ of radius $R$
\beq S_{d}(R)=\Omega_{d}R^{d}=\frac{2\pi^{(d+1)/2}}{\Gamma\left(\frac{d+1}{2}\right)}R^{d}\;,\eeq
and $\mathcal{V}_{d}$ is the (Euclidean) spacetime volume enclosed by $\Sigma_{d}$,
\beq \mathcal{V}_{d}=\frac{2\pi^{(d+1)/2}}{\Gamma\left(\frac{d+1}{2}\right)}\frac{R^{d+1}}{(d+1)}\;.\eeq
In particular, substituting $R$ (\ref{eq:flatrad}) into the on-shell action (\ref{eq:Euconshellflatapp}) we find
\beq
\begin{split}
&S_{E}^{(d=1)}=\frac{\mathcal{M}^{2}\pi}{qE}=\frac{\pi\mathcal{E}_{0}^{2}}{4qE}\;,\\
&S_{E}^{(d=2)}=\frac{16\mathcal{M}^{3}\pi}{3q^{2}E^{2}}=\frac{2}{3}\frac{\mathcal{E}_{0}^{3/2}}{\sqrt{\pi qE}}\;,\\
&S_{E}^{(d=3)}=\frac{27\mathcal{M}^{4}\pi^{2}}{2q^{3}E^{3}}=\left(\frac{3\pi^{2}}{4qE}\right)^{1/3}\frac{\mathcal{E}_{0}^{4/3}}{8}\;.
\end{split}
\label{eq:SEf3at1app}\eeq
When multiplied by an overall winding number $n$, the $d=1$ Euclidean action is precisely the factor which appears in Schwinger's formula for the producton rate of a particle-antiparticle pair in a constant electric field background.

\subsection{Membranes in de Sitter space}

 We now review the nucleation of charged spherical membranes in $(d+1)$-dimensional de Sitter space analyzed in \cite{Garriga:1993fh}. One starts with the action (\ref{eq:lorentactmemflat}), however, where now it is understood the membrane is embedded in the ambient de Sitter background. This will alter the form of the solution and the rest energy relation to the parameter $\mathcal{M}$. Specifically, the Lorentzian solution for a spherical membrane in $(d+1)$-dimensional de Sitter space in static patch coordinates is
\beq r(t)=\frac{\text{sech}(Ht)}{H}\sqrt{\sinh^{2}(Ht)+H^{2}R^{2}}\;,\eeq
such that the energy $\mathcal{E}$ is
 \beq \mathcal{E}=\Pi_{t}=\mathcal{M}\Omega_{d-1}r^{d-1}\frac{\left(1-H^{2}r^{2}\right)^{3/2}}{\sqrt{(1-H^{2}r^{2})^{2}-\dot{r}^{2}}}\;.\eeq
 Consequently, the rest energy $\mathcal{E}_{0}$ is
 \beq \mathcal{E}_{0}=\mathcal{M}\Omega_{d-1}R^{d-1}\sqrt{1-H^{2}R^{2}}\;,\label{eq:E0desitterapp}\eeq
 consistent with (\ref{eq:Enmemflatapp}) in the flat space limit $H\to0$.

Upon Euclideanization, spherical worldvolumes were found to be extrema of the Euclidean action (\ref{eq:IEmemgarriv2}) such that the on-shell action for spherical membranes in de Sitter space is \cite{Garriga:1993fh}
\beq S_{E}=\mathcal{M}S_{d}(R)-qE\mathcal{V}_{d}(\alpha)\;.\label{eq:sphericalmemactapp}\eeq
This is nearly identical to the flat space case (\ref{eq:Euconshellflatapp}), however, now the worldvolume is of radius $R=H^{-1}\sin\alpha$, with $H^{-1}$ being the radius of the $(d+1)$-dimensional  sphere found upon Wick rotating Lorentzian de Sitter, and $\alpha$ its polar angle, and, further, $\mathcal{V}_{d}$ is the volume
\beq \mathcal{V}_{d}(\alpha)=\Omega_{d}H^{-(d+1)}\int_{0}^{\alpha}d\theta \sin^{d}\theta\;.\eeq
The volume $\mathcal{V}_{d}$ may be cast in terms of hypergeometric functions for any positive integer $d$.

Extremizing the on-shell action (\ref{eq:sphericalmemactapp}) with respect to $\alpha$ fixes the radius $R$ to be
\beq R=\frac{d\mathcal{M}}{\sqrt{d^{2}H^{2}\mathcal{M}^{2}+q^{2}E^{2}}}=\left(\frac{\mathcal{E}_{0}d}{qE\Omega_{d-1}}\right)^{1/d}\;,\label{eq:R0garr}\eeq
where the point out the second equality is equivalent to the flat space radius (\ref{eq:flatrad}). From the first equality it is easy to verify $qE\tan(\alpha)=d\mathcal{M}H$. Further, when $H\to0$ one recovers the circular worldline solution of radius $R=\mathcal{M}/qE$ in $(1+1)$-dimensions, which is exactly equal to the distance between two charged particles whose electric potential energy balances their rest mass. Substituting $R$ (\ref{eq:R0garr}) into the on-shell action (\ref{eq:sphericalmemactapp}) yields, for example,
\beq
\begin{split}
&S_{E}^{(d=1)}=\frac{2\pi}{H^{2}}\left[\sqrt{\mathcal{M}^{2}H^{2}+q^{2}E^{2}}-qE\right]\;,\\
&S_{E}^{(d=2)}=\frac{4\pi}{H^{3}}\left[\mathcal{M}H-\frac{qE}{2}\text{arctan}\left(\frac{2 H\mathcal{M}}{qE}\right)\right]\;,\\
&S_{E}^{(d=3)}=\frac{2\pi^{2}}{3H^{4}}\left[\frac{9H^{2}\mathcal{M}^{2}+2q^{2}E^{2}}{(9H^{2}\mathcal{M}^{2}+q^{2}E^{2})^{1/2}}-2qE\right]\;.
\end{split}
\eeq
We can easily reexpress the on-shell action in terms of $\mathcal{E}_{0}$, however, it is generally more cumbersome. For  example, for $d=1$, substituting (\ref{eq:E0desitterapp}) and (\ref{eq:R0garr}) in for $\mathcal{M}$ leads to
\beq S_{E}^{(d=1)}=\frac{2\pi}{H^{2}}\frac{qE}{\sqrt{1-\left(\frac{\mathcal{E}_{0}H}{2qE}\right)^{2}}}\left(1-\sqrt{1-\left(\frac{\mathcal{E}_{0}H}{2qE}\right)^{2}}\right)\;.\label{eq:dSN1E0app}\eeq
 Regardless, notice that in the flat space limit, for the screening configuration\footnote{The flat space limit has infinite action for the anti-screening configuration.} ($q>0$), the Euclidean action is finite, and is consistent with (\ref{eq:SEf3at1app}). Finally, in the limit $E\to0$, such that the coupling to the antisymmetric tensor field is turned off and $R=H^{-1}$, the on-shell action remains finite (contrasting the flat space limit), and one recovers the instanton solutions describing spontaneous nucleation of defects during inflation \cite{Basu:1991ig,Basu:1992ue}.

%%%%%%%%%%%%%%%%%%%%%%%%%%%%%%%%%%%%%%%%%%
\section{De Sitter foliations of Anti de Sitter space} \label{app:dSfoliationsAdS}

Here we list several de Sitter ($\text{dS}_{d+1}$) foliations of $\text{AdS}_{d+2}$. Specifically, following \cite{Fischler:2013fba,Fischler:2014tka,Fischler:2014ama}, we write the $\text{AdS}_{d+2}$ line element as
\beq ds^{2}_{d+2}=\frac{L^{2}}{z^{2}}\left[\left(1-\frac{H^{2}z^{2}}{4}\right)^{2}ds^{2}_{\text{dS}}+dz^{2}\right]\;,\label{eq:AdSlineeledSapp}\eeq
where $ds^{2}_{\text{dS}}$ is a $(d+1)$-dimensional line element for de Sitter in \emph{any} coordinate system, with $H$ being the Hubble constant. Redefining the bulk $z$ coordinate via
\beq z=\frac{2}{H}e^{-2\text{arctanh}(u)}=\frac{2}{H}\frac{(1-u)}{(1+u)}\;,\eeq
where $u\in[0,1]$, with $u=0$ corresponding to the acceleration horizon ($z=2/H$) and $u=1$ the conformal boundary of AdS ($z=0$), leads to
\beq ds^{2}_{d+2}=\frac{4L^{2}}{(1-u^{2})^{2}}\left[H^{2}u^{2}ds_{\text{dS}}^{2}+du^{2}\right]\;.\label{eq:AdSmetdSapp}\eeq
Let us now explore such foliations of $\text{AdS}_{d+2}$ due to various coordinate frames of $\text{dS}_{d+1}$.

\subsection*{Global foliations}

First consider when the AdS metric is foliated by dS in global coordinates,
\beq ds^{2}_{\text{dS}}=-d\tau^{2}+\frac{1}{H^{2}}\cosh^{2}(H\tau)d\Omega_{d}^{2}\;,\eeq
where $\tau$ is time coordinate and $d\Omega_{d}^{2}$ is the line element for the $d$-dimensional sphere; explicitly,
\beq d\Omega_{d}^{2}=d\phi^{2}+\sin\phi^{2} d\psi_{1}^{2}+\sin^{2}\phi\sin^{2}\psi_{1} d\psi_{2}^{2}+...=d\phi^{2}+\sin^{2}\phi d\Omega_{d-1}^{2}\;.\eeq
In the context of cosmology, dS in global coordinates is sometimes called the `closed slicing' as it represents a closed FRW cosmology for an isotropic, homogeneous universe, with a scale factor $a(\tau)=H^{-1}\cosh(H\tau)$.

The Euclidean dS spacetime, which is achieved by Wick rotating $\tau\to -i\tau_{E}/H$ (such that $\tau_{E}$ is dimensionless), yields
\beq ds^{2}_{\text{dS}}=\frac{1}{H^{2}}d\tau_{E}^{2}+\frac{1}{H^{2}}\cos^{2}(\tau_{E})(d\phi^{2}+\sin^{2}\phi d\Omega_{d-1}^{2})\;.\label{eq:dSglobalapp}\eeq
With the Euclidean coordinate system (\ref{eq:dSglobalapp}), the Euclidean $\text{AdS}_{d+2}$ line element is
\beq ds^{2}_{d+2}=\frac{4L^{2}}{(1-u^{2})^{2}}\biggr(du^{2}+u^{2}[d\tau_{E}^{2}+\cos^{2}(\tau_{E})(d\phi^{2}+\sin^{2}\phi d\Omega_{d-1}^{2})]\biggr)\;.\label{eq:Eucballcoordapp}\eeq
This line element for Euclidean AdS is known as the Poincar\'e ball, where the boundary is a $S^{d+1}$ sphere, and $\tau_{E}$ is a polar angle.

\subsection*{Static patch foliations}

In static patch coordinates, one commonly writes the $\text{dS}_{d+1}$ line element as
\beq ds^{2}_{\text{dS}}=-\left(1-H^{2}r^{2}\right)dt^{2}+\left(1-H^{2}r^{2}\right)^{-1}dr^{2}+r^{2}d\Omega_{d-1}^{2}\;,\eeq
for radial coordinate $0<r<H^{-1}$. The transformation which moves us from global coordinates to static patch coordinates is
\beq \tau=H^{-1}\text{arcsinh}\left[-\sqrt{1-H^{2}r^{2}}\sinh(Ht)\right]\;,\quad \phi=\text{arctan}\left[-\frac{Hr}{\sqrt{1-H^{2}r^{2}}}\text{sech}(Ht)\right]\;.\label{eq:coordchglobstat}\eeq
The static patch line element manifestly has a timelike Killing symmetry $\partial_{t}$, and describes the geometry of a single geodesic observer in causal contact with only a portion of the full de Sitter geometry, due to the presence of a cosmological horizon $r=H^{-1}$. The cosmological horizon behaves similar to a black hole horizon, having a Gibbons-Hawking temperature $T_{\text{dS}}=\frac{H}{2\pi}$. Since this foliation only covers a portion of the full spacetime geometry, such that with respect to this foliation one only describes the hyperbolic patch of $\text{AdS}_{2}$, in which the Killing horizon $z=2/H$ is identified with a Rindler horizon.\footnote{Consequently, a pure state from the point of view global AdS will appear mixed due to entanglement between degrees of freedom inside and outside of the acceleration, providing a bulk description of thermality in the boundary (see, \emph{e.g.}, \cite{Chowdhury:2014oba}.}

It is useful to introduce an angular coordinate $\theta$
\beq Hr=\sin\theta\;,\eeq
for $0<\theta<\pi/2$, where the cosmological horizon appears at $\theta=\pi/2$. Then, the static patch line element becomes
\beq ds^{2}_{\text{dS}}=-\cos^{2}\theta dt^{2}+\frac{1}{H^{2}}\left(d\theta^{2}+\sin^{2}\theta d\Omega_{d-1}^{2}\right)\;.\eeq
Wick rotating such that $t\to -it_{E}/H$, we have Euclidean dS in static patch coordinates become
\beq ds^{2}_{\text{dS}}=\frac{1}{H^{2}}\left(\cos^{2}\theta dt_{E}^{2}+d\theta^{2}+\sin^{2}\theta d\Omega_{d-1}^{2}\right)\;.\label{eq:dSstatEucapp}\eeq
To avoid a conical singularity at $\theta=\pi/2$, one periodically identifies the Euclidean time coordinate, $t_{E}\sim t_{E}+2\pi$, an azimuthal angle.  Cast in this way, the line element is manifestly a $S^{d+1}$ sphere in Hopf-like coordinates (cf. \cite{Martin:2020api}).

Substituting (\ref{eq:dSstatEucapp}) into the AdS metric (\ref{eq:AdSmetdSapp}) yields
\beq ds^{2}_{d+2}=\frac{4L^{2}}{(1-u^{2})^{2}}[du^{2}+u^{2}[\cos^{2}\theta dt_{E}^{2}+d\theta^{2}+\sin^{2}\theta d\Omega_{d-1}^{2}]\;.\label{eq:AdSEucstatapp}\eeq
Since the static patch observer in Lorentzian coordinates only has access to a portion of the full de Sitter geometry, this should be reflected in the Euclidean geometry (\ref{eq:AdSEucstatapp}). Indeed, the the topology of the Euclidean static patch is a hemisphere as opposed to the whole sphere, such that each static patch observer has a different analytic continuation which amounts to rotating the hemipshere around the sphere.

\subsection*{Poincar\'e cylinder}

Another useful set of coordinates is to express Euclidean AdS as a solid cylinder. To accomplish this one performs the following coordinate transformation from $\text{AdS}_{d+2}$ foliated in static patch coordinates\footnote{Note that, by swapping $t_{E}\leftrightarrow\theta$, one attains the transformation from Euclidean $\text{AdS}_{d+2}$ foliated by global dS to the Poincar\'e cylinder \cite{Krtous:2014pva}.}
\beq
\begin{split}
&P=\frac{2u}{1-u^{2}}\sqrt{1-\cos^{2}t_{E}\cos^{2}\theta}\;,\\
&Z=\text{arccosh}\left(\frac{1+u^{2}}{\sqrt{(1+u^{2})^{2}-4u^{2}\cos^{2}t_{E}\cos^{2}\theta}}\right)\;,\\
&\varphi=\text{arccos}\left(\frac{\cos\theta\sin t_{E}}{\sqrt{1-\cos^{2}t_{E}\cos^{2}\theta}}\right)\;.
\end{split}
\label{eq:PZphicoordapp}\eeq
Thus, the Euclidean $AdS_{d+2}$ line element (\ref{eq:AdSEucstatapp}) becomes
\beq ds^{2}_{d+2}=L^{2}\left[\frac{dP^{2}}{1+P^{2}}+(1+P^{2})dZ^{2}+P^{2}d\Omega_{d}^{2}\right]\;,\eeq
where now
\beq d\Omega_{d}^{2}=d\varphi^{2}+\sin^{2}\varphi d\Omega_{d-1}^{2}\;.\eeq
In these coordinates the boundary of the cylinder appears at $P\to\infty$ ($u=1$), and $\varphi\in[0,\pi]$.

\bibliography{membranesrefs}

\providecommand{\href}[2]{#2}\begingroup\raggedright\begin{thebibliography}{100}

\bibitem{Coleman:1977py}
S.~R. Coleman, \emph{{The Fate of the False Vacuum. 1. Semiclassical Theory}},
  \href{http://dx.doi.org/10.1103/PhysRevD.16.1248}{\emph{Phys. Rev. D} {\bf
  15} (1977) 2929--2936}.

\bibitem{Callan:1977pt}
C.~G. Callan, Jr. and S.~R. Coleman, \emph{{The Fate of the False Vacuum. 2.
  First Quantum Corrections}},
  \href{http://dx.doi.org/10.1103/PhysRevD.16.1762}{\emph{Phys. Rev. D} {\bf
  16} (1977) 1762--1768}.

\bibitem{Coleman:1980aw}
S.~R. Coleman and F.~De~Luccia, \emph{{Gravitational Effects on and of Vacuum
  Decay}}, \href{http://dx.doi.org/10.1103/PhysRevD.21.3305}{\emph{Phys. Rev.
  D} {\bf 21} (1980) 3305}.

\bibitem{Lee:1987qc}
K.-M. Lee and E.~J. Weinberg, \emph{{Decay of the True Vacuum in Curved
  Space-time}}, \href{http://dx.doi.org/10.1103/PhysRevD.36.1088}{\emph{Phys.
  Rev. D} {\bf 36} (1987) 1088}.

\bibitem{Parikh:1999mf}
M.~K. Parikh and F.~Wilczek, \emph{{Hawking radiation as tunneling}},
  \href{http://dx.doi.org/10.1103/PhysRevLett.85.5042}{\emph{Phys. Rev. Lett.}
  {\bf 85} (2000) 5042--5045},
  [\href{https://arxiv.org/abs/hep-th/9907001}{{\tt hep-th/9907001}}].

\bibitem{Hiscock:1987hn}
W.~A. Hiscock, \emph{{CAN BLACK HOLES NUCLEATE VACUUM PHASE TRANSITIONS?}},
  \href{http://dx.doi.org/10.1103/PhysRevD.35.1161}{\emph{Phys. Rev. D} {\bf
  35} (1987) 1161--1170}.

\bibitem{Gregory:2013hja}
R.~Gregory, I.~G. Moss and B.~Withers, \emph{{Black holes as bubble nucleation
  sites}}, \href{http://dx.doi.org/10.1007/JHEP03(2014)081}{\emph{JHEP} {\bf
  03} (2014) 081}, [\href{https://arxiv.org/abs/1401.0017}{{\tt 1401.0017}}].

\bibitem{Burda:2015isa}
P.~Burda, R.~Gregory and I.~Moss, \emph{{Gravity and the stability of the Higgs
  vacuum}}, \href{http://dx.doi.org/10.1103/PhysRevLett.115.071303}{\emph{Phys.
  Rev. Lett.} {\bf 115} (2015) 071303},
  [\href{https://arxiv.org/abs/1501.04937}{{\tt 1501.04937}}].

\bibitem{Schwinger:1951nm}
J.~S. Schwinger, \emph{{On gauge invariance and vacuum polarization}},
  \href{http://dx.doi.org/10.1103/PhysRev.82.664}{\emph{Phys. Rev.} {\bf 82}
  (1951) 664--679}.

\bibitem{Brown:1987dd}
J.~D. Brown and C.~Teitelboim, \emph{{Dynamical Neutralization of the
  Cosmological Constant}},
  \href{http://dx.doi.org/10.1016/0370-2693(87)91190-7}{\emph{Phys. Lett. B}
  {\bf 195} (1987) 177--182}.

\bibitem{Brown:1988kg}
J.~D. Brown and C.~Teitelboim, \emph{{Neutralization of the Cosmological
  Constant by Membrane Creation}},
  \href{http://dx.doi.org/10.1016/0550-3213(88)90559-7}{\emph{Nucl. Phys. B}
  {\bf 297} (1988) 787--836}.

\bibitem{Kaloper:2022oqv}
N.~Kaloper, \emph{{Hidden Variables of Gravity and Geometry and the
  Cosmological Constant Problem}},
  \href{https://arxiv.org/abs/2202.06977}{{\tt 2202.06977}}.

\bibitem{Kaloper:2022utc}
N.~Kaloper, \emph{{Pancosmic Relativity and Nature's Hierarchies}},
  \href{https://arxiv.org/abs/2202.08860}{{\tt 2202.08860}}.

\bibitem{Kaloper:2022jpv}
N.~Kaloper and A.~Westphal, \emph{{A Quantum-Mechanical Mechanism for Reducing
  the Cosmological Constant}},  \href{https://arxiv.org/abs/2204.13124}{{\tt
  2204.13124}}.

\bibitem{Garriga:1993fh}
J.~Garriga, \emph{{Nucleation rates in flat and curved space}},
  \href{http://dx.doi.org/10.1103/PhysRevD.49.6327}{\emph{Phys. Rev. D} {\bf
  49} (1994) 6327--6342}, [\href{https://arxiv.org/abs/hep-ph/9308280}{{\tt
  hep-ph/9308280}}].

\bibitem{Coleman:1985rnk}
S.~Coleman, \emph{{Aspects of Symmetry}: {Selected Erice Lectures}}.
\newblock Cambridge University Press, Cambridge, U.K., 1985,
  \href{http://dx.doi.org/10.1017/CBO9780511565045}{10.1017/CBO9780511565045}.

\bibitem{Affleck:1981bma}
I.~K. Affleck, O.~Alvarez and N.~S. Manton, \emph{{Pair Production at Strong
  Coupling in Weak External Fields}},
  \href{http://dx.doi.org/10.1016/0550-3213(82)90455-2}{\emph{Nucl. Phys. B}
  {\bf 197} (1982) 509--519}.

\bibitem{Basu:1991ig}
R.~Basu, A.~H. Guth and A.~Vilenkin, \emph{{Quantum creation of topological
  defects during inflation}},
  \href{http://dx.doi.org/10.1103/PhysRevD.44.340}{\emph{Phys. Rev. D} {\bf 44}
  (1991) 340--351}.

\bibitem{Garriga:1994bm}
J.~Garriga, \emph{{Pair production by an electric field in (1+1)-dimensional de
  Sitter space}}, \href{http://dx.doi.org/10.1103/PhysRevD.49.6343}{\emph{Phys.
  Rev. D} {\bf 49} (1994) 6343--6346}.

\bibitem{Ryu:2006bv}
S.~Ryu and T.~Takayanagi, \emph{{Holographic derivation of entanglement entropy
  from AdS/CFT}},
  \href{http://dx.doi.org/10.1103/PhysRevLett.96.181602}{\emph{Phys. Rev.
  Lett.} {\bf 96} (2006) 181602},
  [\href{https://arxiv.org/abs/hep-th/0603001}{{\tt hep-th/0603001}}].

\bibitem{Kovtun:2004de}
P.~Kovtun, D.~T. Son and A.~O. Starinets, \emph{{Viscosity in strongly
  interacting quantum field theories from black hole physics}},
  \href{http://dx.doi.org/10.1103/PhysRevLett.94.111601}{\emph{Phys. Rev.
  Lett.} {\bf 94} (2005) 111601},
  [\href{https://arxiv.org/abs/hep-th/0405231}{{\tt hep-th/0405231}}].

\bibitem{Luzum:2008cw}
M.~Luzum and P.~Romatschke, \emph{{Conformal Relativistic Viscous
  Hydrodynamics: Applications to RHIC results at s(NN)**(1/2) = 200-GeV}},
  \href{http://dx.doi.org/10.1103/PhysRevC.78.034915}{\emph{Phys. Rev. C} {\bf
  78} (2008) 034915}, [\href{https://arxiv.org/abs/0804.4015}{{\tt
  0804.4015}}].

\bibitem{Semenoff:2011ng}
G.~W. Semenoff and K.~Zarembo, \emph{{Holographic Schwinger Effect}},
  \href{http://dx.doi.org/10.1103/PhysRevLett.107.171601}{\emph{Phys. Rev.
  Lett.} {\bf 107} (2011) 171601}, [\href{https://arxiv.org/abs/1109.2920}{{\tt
  1109.2920}}].

\bibitem{Ambjorn:2011wz}
J.~Ambjorn and Y.~Makeenko, \emph{{Remarks on Holographic Wilson Loops and the
  Schwinger Effect}},
  \href{http://dx.doi.org/10.1103/PhysRevD.85.061901}{\emph{Phys. Rev. D} {\bf
  85} (2012) 061901}, [\href{https://arxiv.org/abs/1112.5606}{{\tt
  1112.5606}}].

\bibitem{Sato:2013pxa}
Y.~Sato and K.~Yoshida, \emph{{Holographic description of the Schwinger effect
  in electric and magnetic fields}},
  \href{http://dx.doi.org/10.1007/JHEP04(2013)111}{\emph{JHEP} {\bf 04} (2013)
  111}, [\href{https://arxiv.org/abs/1303.0112}{{\tt 1303.0112}}].

\bibitem{Sato:2013iua}
Y.~Sato and K.~Yoshida, \emph{{Potential Analysis in Holographic Schwinger
  Effect}}, \href{http://dx.doi.org/10.1007/JHEP08(2013)002}{\emph{JHEP} {\bf
  08} (2013) 002}, [\href{https://arxiv.org/abs/1304.7917}{{\tt 1304.7917}}].

\bibitem{Kawai:2013xya}
D.~Kawai, Y.~Sato and K.~Yoshida, \emph{{Schwinger pair production rate in
  confining theories via holography}},
  \href{http://dx.doi.org/10.1103/PhysRevD.89.101901}{\emph{Phys. Rev. D} {\bf
  89} (2014) 101901}, [\href{https://arxiv.org/abs/1312.4341}{{\tt
  1312.4341}}].

\bibitem{Wu:2015krf}
X.~Wu, \emph{{Notes on holographic Schwinger effect}},
  \href{http://dx.doi.org/10.1007/JHEP09(2015)044}{\emph{JHEP} {\bf 09} (2015)
  044}, [\href{https://arxiv.org/abs/1507.03208}{{\tt 1507.03208}}].

\bibitem{Sato:2013hyw}
Y.~Sato and K.~Yoshida, \emph{{Universal aspects of holographic Schwinger
  effect in general backgrounds}},
  \href{http://dx.doi.org/10.1007/JHEP12(2013)051}{\emph{JHEP} {\bf 12} (2013)
  051}, [\href{https://arxiv.org/abs/1309.4629}{{\tt 1309.4629}}].

\bibitem{Fischler:2014ama}
W.~Fischler, P.~H. Nguyen, J.~F. Pedraza and W.~Tangarife, \emph{{Holographic
  Schwinger effect in de Sitter space}},
  \href{http://dx.doi.org/10.1103/PhysRevD.91.086015}{\emph{Phys. Rev. D} {\bf
  91} (2015) 086015}, [\href{https://arxiv.org/abs/1411.1787}{{\tt
  1411.1787}}].

\bibitem{BitaghsirFadafan:2015asm}
K.~Bitaghsir~Fadafan and F.~Saiedi, \emph{{Holographic Schwinger effect in
  non-relativistic backgrounds}},
  \href{http://dx.doi.org/10.1140/epjc/s10052-015-3839-1}{\emph{Eur. Phys. J.
  C} {\bf 75} (2015) 612}, [\href{https://arxiv.org/abs/1504.02432}{{\tt
  1504.02432}}].

\bibitem{Kawai:2015mha}
D.~Kawai, Y.~Sato and K.~Yoshida, \emph{{A holographic description of the
  Schwinger effect in a confining gauge theory}},
  \href{http://dx.doi.org/10.1142/S0217751X15300264}{\emph{Int. J. Mod. Phys.
  A} {\bf 30} (2015) 1530026}, [\href{https://arxiv.org/abs/1504.00459}{{\tt
  1504.00459}}].

\bibitem{Ghodrati:2015rta}
M.~Ghodrati, \emph{{Schwinger Effect and Entanglement Entropy in Confining
  Geometries}}, \href{http://dx.doi.org/10.1103/PhysRevD.92.065015}{\emph{Phys.
  Rev. D} {\bf 92} (2015) 065015},
  [\href{https://arxiv.org/abs/1506.08557}{{\tt 1506.08557}}].

\bibitem{Shahkarami:2015qff}
L.~Shahkarami, M.~Dehghani and P.~Dehghani, \emph{{Holographic Schwinger Effect
  in a D-Instanton Background}},
  \href{http://dx.doi.org/10.1103/PhysRevD.97.046013}{\emph{Phys. Rev. D} {\bf
  97} (2018) 046013}, [\href{https://arxiv.org/abs/1511.07986}{{\tt
  1511.07986}}].

\bibitem{Villalba:1995za}
V.~M. Villalba, \emph{{Creation of spin 1/2 particles by an electric field in
  de Sitter space}},
  \href{http://dx.doi.org/10.1103/PhysRevD.52.3742}{\emph{Phys. Rev. D} {\bf
  52} (1995) 3742--3745}, [\href{https://arxiv.org/abs/hep-th/9507021}{{\tt
  hep-th/9507021}}].

\bibitem{Kim:2008xv}
S.~P. Kim and D.~N. Page, \emph{{Schwinger Pair Production in dS(2) and
  AdS(2)}}, \href{http://dx.doi.org/10.1103/PhysRevD.78.103517}{\emph{Phys.
  Rev. D} {\bf 78} (2008) 103517}, [\href{https://arxiv.org/abs/0803.2555}{{\tt
  0803.2555}}].

\bibitem{Stahl:2015gaa}
C.~Stahl, E.~Strobel and S.-S. Xue, \emph{{Fermionic current and Schwinger
  effect in de Sitter spacetime}},
  \href{http://dx.doi.org/10.1103/PhysRevD.93.025004}{\emph{Phys. Rev. D} {\bf
  93} (2016) 025004}, [\href{https://arxiv.org/abs/1507.01686}{{\tt
  1507.01686}}].

\bibitem{Bavarsad:2016cxh}
E.~Bavarsad, C.~Stahl and S.-S. Xue, \emph{{Scalar current of created pairs by
  Schwinger mechanism in de Sitter spacetime}},
  \href{http://dx.doi.org/10.1103/PhysRevD.94.104011}{\emph{Phys. Rev. D} {\bf
  94} (2016) 104011}, [\href{https://arxiv.org/abs/1602.06556}{{\tt
  1602.06556}}].

\bibitem{Stahl:2016geq}
C.~Stahl and S.-S. Xue, \emph{{Schwinger effect and backreaction in de Sitter
  spacetime}},
  \href{http://dx.doi.org/10.1016/j.physletb.2016.07.011}{\emph{Phys. Lett. B}
  {\bf 760} (2016) 288--292}, [\href{https://arxiv.org/abs/1603.07166}{{\tt
  1603.07166}}].

\bibitem{Sharma:2017ivh}
R.~Sharma and S.~Singh, \emph{{Multifaceted Schwinger effect in de Sitter
  space}}, \href{http://dx.doi.org/10.1103/PhysRevD.96.025012}{\emph{Phys. Rev.
  D} {\bf 96} (2017) 025012}, [\href{https://arxiv.org/abs/1704.05076}{{\tt
  1704.05076}}].

\bibitem{Grewal:2021bsu}
M.~Grewal and K.~Parmentier, \emph{{Characters, quasinormal modes, and
  Schwinger pairs in dS$_{2}$ with flux}},
  \href{http://dx.doi.org/10.1007/JHEP03(2022)165}{\emph{JHEP} {\bf 03} (2022)
  165}, [\href{https://arxiv.org/abs/2112.07630}{{\tt 2112.07630}}].

\bibitem{Gubser:1998bc}
S.~S. Gubser, I.~R. Klebanov and A.~M. Polyakov, \emph{{Gauge theory
  correlators from noncritical string theory}},
  \href{http://dx.doi.org/10.1016/S0370-2693(98)00377-3}{\emph{Phys. Lett. B}
  {\bf 428} (1998) 105--114}, [\href{https://arxiv.org/abs/hep-th/9802109}{{\tt
  hep-th/9802109}}].

\bibitem{Witten:1998qj}
E.~Witten, \emph{{Anti-de Sitter space and holography}},
  \href{http://dx.doi.org/10.4310/ATMP.1998.v2.n2.a2}{\emph{Adv. Theor. Math.
  Phys.} {\bf 2} (1998) 253--291},
  [\href{https://arxiv.org/abs/hep-th/9802150}{{\tt hep-th/9802150}}].

\bibitem{Karch:2002sh}
A.~Karch and E.~Katz, \emph{{Adding flavor to AdS / CFT}},
  \href{http://dx.doi.org/10.1088/1126-6708/2002/06/043}{\emph{JHEP} {\bf 06}
  (2002) 043}, [\href{https://arxiv.org/abs/hep-th/0205236}{{\tt
  hep-th/0205236}}].

\bibitem{Randall:1999vf}
L.~Randall and R.~Sundrum, \emph{{An Alternative to compactification}},
  \href{http://dx.doi.org/10.1103/PhysRevLett.83.4690}{\emph{Phys. Rev. Lett.}
  {\bf 83} (1999) 4690--4693},
  [\href{https://arxiv.org/abs/hep-th/9906064}{{\tt hep-th/9906064}}].

\bibitem{Randall:1999ee}
L.~Randall and R.~Sundrum, \emph{{A Large mass hierarchy from a small extra
  dimension}}, \href{http://dx.doi.org/10.1103/PhysRevLett.83.3370}{\emph{Phys.
  Rev. Lett.} {\bf 83} (1999) 3370--3373},
  [\href{https://arxiv.org/abs/hep-ph/9905221}{{\tt hep-ph/9905221}}].

\bibitem{Emparan:2002px}
R.~Emparan, A.~Fabbri and N.~Kaloper, \emph{{Quantum black holes as holograms
  in AdS brane worlds}},
  \href{http://dx.doi.org/10.1088/1126-6708/2002/08/043}{\emph{JHEP} {\bf 08}
  (2002) 043}, [\href{https://arxiv.org/abs/hep-th/0206155}{{\tt
  hep-th/0206155}}].

\bibitem{Emparan:2020znc}
R.~Emparan, A.~M. Frassino and B.~Way, \emph{{Quantum BTZ black hole}},
  \href{http://dx.doi.org/10.1007/JHEP11(2020)137}{\emph{JHEP} {\bf 11} (2020)
  137}, [\href{https://arxiv.org/abs/2007.15999}{{\tt 2007.15999}}].

\bibitem{Emparan:2022ijy}
R.~Emparan, J.~F. Pedraza, A.~Svesko, M.~Toma\v{s}evi\'c and M.~R. Visser,
  \emph{{Black holes in dS$_3$}},  \href{https://arxiv.org/abs/2207.03302}{{\tt
  2207.03302}}.

\bibitem{Xiao:2008nr}
B.-W. Xiao, \emph{{On the exact solution of the accelerating string in AdS(5)
  space}}, \href{http://dx.doi.org/10.1016/j.physletb.2008.06.017}{\emph{Phys.
  Lett. B} {\bf 665} (2008) 173--177},
  [\href{https://arxiv.org/abs/0804.1343}{{\tt 0804.1343}}].

\bibitem{Caceres:2010rm}
E.~Caceres, M.~Chernicoff, A.~Guijosa and J.~F. Pedraza, \emph{{Quantum
  Fluctuations and the Unruh Effect in Strongly-Coupled Conformal Field
  Theories}}, \href{http://dx.doi.org/10.1007/JHEP06(2010)078}{\emph{JHEP} {\bf
  06} (2010) 078}, [\href{https://arxiv.org/abs/1003.5332}{{\tt 1003.5332}}].

\bibitem{Hovdebo:2005hm}
J.~L. Hovdebo, M.~Kruczenski, D.~Mateos, R.~C. Myers and D.~J. Winters,
  \emph{{Holographic mesons: Adding flavor to the AdS/CFT duality}},
  \href{http://dx.doi.org/10.1142/S0217751X05026728}{\emph{Int. J. Mod. Phys.
  A} {\bf 20} (2005) 3428--3433}.

\bibitem{Chernicoff:2009re}
M.~Chernicoff, J.~A. Garcia and A.~Guijosa, \emph{{Generalized Lorentz-Dirac
  Equation for a Strongly-Coupled Gauge Theory}},
  \href{http://dx.doi.org/10.1103/PhysRevLett.102.241601}{\emph{Phys. Rev.
  Lett.} {\bf 102} (2009) 241601}, [\href{https://arxiv.org/abs/0903.2047}{{\tt
  0903.2047}}].

\bibitem{Chernicoff:2009ff}
M.~Chernicoff, J.~A. Garcia and A.~Guijosa, \emph{{A Tail of a Quark in N=4
  SYM}}, \href{http://dx.doi.org/10.1088/1126-6708/2009/09/080}{\emph{JHEP}
  {\bf 09} (2009) 080}, [\href{https://arxiv.org/abs/0906.1592}{{\tt
  0906.1592}}].

\bibitem{Chernicoff:2011vn}
M.~Chernicoff, A.~Guijosa and J.~F. Pedraza, \emph{{The Gluonic Field of a
  Heavy Quark in Conformal Field Theories at Strong Coupling}},
  \href{http://dx.doi.org/10.1007/JHEP10(2011)041}{\emph{JHEP} {\bf 10} (2011)
  041}, [\href{https://arxiv.org/abs/1106.4059}{{\tt 1106.4059}}].

\bibitem{Agon:2014rda}
C.~A. Ag\'on, A.~Guijosa and J.~F. Pedraza, \emph{{Radiation and a dynamical
  UV/IR connection in AdS/CFT}},
  \href{http://dx.doi.org/10.1007/JHEP06(2014)043}{\emph{JHEP} {\bf 06} (2014)
  043}, [\href{https://arxiv.org/abs/1402.5961}{{\tt 1402.5961}}].

\bibitem{Chernicoff:2011xv}
M.~Chernicoff, J.~A. Garcia, A.~Guijosa and J.~F. Pedraza, \emph{{Holographic
  Lessons for Quark Dynamics}},
  \href{http://dx.doi.org/10.1088/0954-3899/39/5/054002}{\emph{J. Phys. G} {\bf
  39} (2012) 054002}, [\href{https://arxiv.org/abs/1111.0872}{{\tt
  1111.0872}}].

\bibitem{Cohen:2008wz}
T.~D. Cohen and D.~A. McGady, \emph{{The Schwinger mechanism revisited}},
  \href{http://dx.doi.org/10.1103/PhysRevD.78.036008}{\emph{Phys. Rev. D} {\bf
  78} (2008) 036008}, [\href{https://arxiv.org/abs/0807.1117}{{\tt
  0807.1117}}].

\bibitem{Marolf:2013ioa}
D.~Marolf, M.~Rangamani and T.~Wiseman, \emph{{Holographic thermal field theory
  on curved spacetimes}},
  \href{http://dx.doi.org/10.1088/0264-9381/31/6/063001}{\emph{Class. Quant.
  Grav.} {\bf 31} (2014) 063001}, [\href{https://arxiv.org/abs/1312.0612}{{\tt
  1312.0612}}].

\bibitem{Fischler:2013fba}
W.~Fischler, S.~Kundu and J.~F. Pedraza, \emph{{Entanglement and
  out-of-equilibrium dynamics in holographic models of de Sitter QFTs}},
  \href{http://dx.doi.org/10.1007/JHEP07(2014)021}{\emph{JHEP} {\bf 07} (2014)
  021}, [\href{https://arxiv.org/abs/1311.5519}{{\tt 1311.5519}}].

\bibitem{Fischler:2014tka}
W.~Fischler, P.~H. Nguyen, J.~F. Pedraza and W.~Tangarife, \emph{{Fluctuation
  and dissipation in de Sitter space}},
  \href{http://dx.doi.org/10.1007/JHEP08(2014)028}{\emph{JHEP} {\bf 08} (2014)
  028}, [\href{https://arxiv.org/abs/1404.0347}{{\tt 1404.0347}}].

\bibitem{Zhang:2014cga}
S.-J. Zhang, B.~Wang, E.~Abdalla and E.~Papantonopoulos, \emph{{Holographic
  thermalization in Gauss-Bonnet gravity with de Sitter boundary}},
  \href{http://dx.doi.org/10.1103/PhysRevD.91.106010}{\emph{Phys. Rev. D} {\bf
  91} (2015) 106010}, [\href{https://arxiv.org/abs/1412.7073}{{\tt
  1412.7073}}].

\bibitem{Chu:2016uwi}
C.-S. Chu and D.~Giataganas, \emph{{AdS/dS CFT Correspondence}},
  \href{http://dx.doi.org/10.1103/PhysRevD.94.106013}{\emph{Phys. Rev. D} {\bf
  94} (2016) 106013}, [\href{https://arxiv.org/abs/1604.05452}{{\tt
  1604.05452}}].

\bibitem{Chu:2016pea}
C.-S. Chu and D.~Giataganas, \emph{{Thermal bath in de Sitter space from
  holography}}, \href{http://dx.doi.org/10.1103/PhysRevD.96.026023}{\emph{Phys.
  Rev. D} {\bf 96} (2017) 026023},
  [\href{https://arxiv.org/abs/1608.07431}{{\tt 1608.07431}}].

\bibitem{Zhang:2019vgl}
S.-J. Zhang, \emph{{Subregion complexity in holographic thermalization with dS
  boundary}},
  \href{http://dx.doi.org/10.1140/epjc/s10052-019-7241-2}{\emph{Eur. Phys. J.
  C} {\bf 79} (2019) 715}, [\href{https://arxiv.org/abs/1905.10605}{{\tt
  1905.10605}}].

\bibitem{Ageev:2021xjk}
D.~S. Ageev, \emph{{Butterfly velocity and chaos suppression in de Sitter
  space}},  \href{https://arxiv.org/abs/2105.02258}{{\tt 2105.02258}}.

\bibitem{Krtous:2014pva}
P.~Krtous and A.~Zelnikov, \emph{{Minimal surfaces and entanglement entropy in
  anti-de Sitter space}},
  \href{http://dx.doi.org/10.1007/JHEP10(2014)077}{\emph{JHEP} {\bf 10} (2014)
  077}, [\href{https://arxiv.org/abs/1406.7659}{{\tt 1406.7659}}].

\bibitem{Basu:1992ue}
R.~Basu and A.~Vilenkin, \emph{{Nucleation of thick topological defects during
  inflation}}, \href{http://dx.doi.org/10.1103/PhysRevD.46.2345}{\emph{Phys.
  Rev. D} {\bf 46} (1992) 2345--2354}.

\bibitem{Guth:1982pn}
A.~H. Guth and E.~J. Weinberg, \emph{{Could the Universe Have Recovered from a
  Slow First Order Phase Transition?}},
  \href{http://dx.doi.org/10.1016/0550-3213(83)90307-3}{\emph{Nucl. Phys. B}
  {\bf 212} (1983) 321--364}.

\bibitem{Garriga:2006hw}
J.~Garriga, A.~H. Guth and A.~Vilenkin, \emph{{Eternal inflation, bubble
  collisions, and the persistence of memory}},
  \href{http://dx.doi.org/10.1103/PhysRevD.76.123512}{\emph{Phys. Rev. D} {\bf
  76} (2007) 123512}, [\href{https://arxiv.org/abs/hep-th/0612242}{{\tt
  hep-th/0612242}}].

\bibitem{Freivogel:2007fx}
B.~Freivogel, G.~T. Horowitz and S.~Shenker, \emph{{Colliding with a crunching
  bubble}}, \href{http://dx.doi.org/10.1088/1126-6708/2007/05/090}{\emph{JHEP}
  {\bf 05} (2007) 090}, [\href{https://arxiv.org/abs/hep-th/0703146}{{\tt
  hep-th/0703146}}].

\bibitem{Chang:2007eq}
S.~Chang, M.~Kleban and T.~S. Levi, \emph{{When worlds collide}},
  \href{http://dx.doi.org/10.1088/1475-7516/2008/04/034}{\emph{JCAP} {\bf 04}
  (2008) 034}, [\href{https://arxiv.org/abs/0712.2261}{{\tt 0712.2261}}].

\bibitem{Aguirre:2007an}
A.~Aguirre, M.~C. Johnson and A.~Shomer, \emph{{Towards observable signatures
  of other bubble universes}},
  \href{http://dx.doi.org/10.1103/PhysRevD.76.063509}{\emph{Phys. Rev. D} {\bf
  76} (2007) 063509}, [\href{https://arxiv.org/abs/0704.3473}{{\tt
  0704.3473}}].

\bibitem{Aguirre:2007wm}
A.~Aguirre and M.~C. Johnson, \emph{{Towards observable signatures of other
  bubble universes. II: Exact solutions for thin-wall bubble collisions}},
  \href{http://dx.doi.org/10.1103/PhysRevD.77.123536}{\emph{Phys. Rev. D} {\bf
  77} (2008) 123536}, [\href{https://arxiv.org/abs/0712.3038}{{\tt
  0712.3038}}].

\bibitem{Chang:2008gj}
S.~Chang, M.~Kleban and T.~S. Levi, \emph{{Watching Worlds Collide: Effects on
  the CMB from Cosmological Bubble Collisions}},
  \href{http://dx.doi.org/10.1088/1475-7516/2009/04/025}{\emph{JCAP} {\bf 04}
  (2009) 025}, [\href{https://arxiv.org/abs/0810.5128}{{\tt 0810.5128}}].

\bibitem{Vegh:2021sva}
D.~Vegh, \emph{{Relativistic membrane solutions in AdS$_4$}},
  \href{https://arxiv.org/abs/2101.03143}{{\tt 2101.03143}}.

\bibitem{Alberghi:1999kd}
G.~L. Alberghi, D.~A. Lowe and M.~Trodden, \emph{{Charged false vacuum bubbles
  and the AdS / CFT correspondence}},
  \href{http://dx.doi.org/10.1088/1126-6708/1999/07/020}{\emph{JHEP} {\bf 07}
  (1999) 020}, [\href{https://arxiv.org/abs/hep-th/9906047}{{\tt
  hep-th/9906047}}].

\bibitem{Balasubramanian:2002am}
V.~Balasubramanian and S.~F. Ross, \emph{{The Dual of nothing}},
  \href{http://dx.doi.org/10.1103/PhysRevD.66.086002}{\emph{Phys. Rev. D} {\bf
  66} (2002) 086002}, [\href{https://arxiv.org/abs/hep-th/0205290}{{\tt
  hep-th/0205290}}].

\bibitem{Banks:2002nm}
T.~Banks, \emph{{Heretics of the false vacuum: Gravitational effects on and of
  vacuum decay. 2.}},  \href{https://arxiv.org/abs/hep-th/0211160}{{\tt
  hep-th/0211160}}.

\bibitem{Ross:2004cb}
S.~F. Ross and G.~Titchener, \emph{{Time-dependent spacetimes in AdS/CFT:
  Bubble and black hole}},
  \href{http://dx.doi.org/10.1088/1126-6708/2005/02/021}{\emph{JHEP} {\bf 02}
  (2005) 021}, [\href{https://arxiv.org/abs/hep-th/0411128}{{\tt
  hep-th/0411128}}].

\bibitem{Balasubramanian:2005bg}
V.~Balasubramanian, K.~Larjo and J.~Simon, \emph{{Much ado about nothing}},
  \href{http://dx.doi.org/10.1088/0264-9381/22/19/023}{\emph{Class. Quant.
  Grav.} {\bf 22} (2005) 4149--4170},
  [\href{https://arxiv.org/abs/hep-th/0502111}{{\tt hep-th/0502111}}].

\bibitem{deHaro:2006ymc}
S.~de~Haro, I.~Papadimitriou and A.~C. Petkou, \emph{{Conformally Coupled
  Scalars, Instantons and Vacuum Instability in AdS(4)}},
  \href{http://dx.doi.org/10.1103/PhysRevLett.98.231601}{\emph{Phys. Rev.
  Lett.} {\bf 98} (2007) 231601},
  [\href{https://arxiv.org/abs/hep-th/0611315}{{\tt hep-th/0611315}}].

\bibitem{Papadimitriou:2007sj}
I.~Papadimitriou, \emph{{Multi-Trace Deformations in AdS/CFT: Exploring the
  Vacuum Structure of the Deformed CFT}},
  \href{http://dx.doi.org/10.1088/1126-6708/2007/05/075}{\emph{JHEP} {\bf 05}
  (2007) 075}, [\href{https://arxiv.org/abs/hep-th/0703152}{{\tt
  hep-th/0703152}}].

\bibitem{He:2007ji}
J.~He and M.~Rozali, \emph{{On bubbles of nothing in AdS/CFT}},
  \href{http://dx.doi.org/10.1088/1126-6708/2007/09/089}{\emph{JHEP} {\bf 09}
  (2007) 089}, [\href{https://arxiv.org/abs/hep-th/0703220}{{\tt
  hep-th/0703220}}].

\bibitem{Barbon:2010gn}
J.~L.~F. Barbon and E.~Rabinovici, \emph{{Holography of AdS vacuum bubbles}},
  \href{http://dx.doi.org/10.1007/JHEP04(2010)123}{\emph{JHEP} {\bf 04} (2010)
  123}, [\href{https://arxiv.org/abs/1003.4966}{{\tt 1003.4966}}].

\bibitem{Maldacena:2010un}
J.~Maldacena, \emph{{Vacuum decay into Anti de Sitter space}},
  \href{https://arxiv.org/abs/1012.0274}{{\tt 1012.0274}}.

\bibitem{Barbon:2011ta}
J.~L.~F. Barbon and E.~Rabinovici, \emph{{AdS Crunches, CFT Falls And
  Cosmological Complementarity}},
  \href{http://dx.doi.org/10.1007/JHEP04(2011)044}{\emph{JHEP} {\bf 04} (2011)
  044}, [\href{https://arxiv.org/abs/1102.3015}{{\tt 1102.3015}}].

\bibitem{Ghosh:2021lua}
J.~K. Ghosh, E.~Kiritsis, F.~Nitti and L.~T. Witkowski, \emph{{Revisiting
  Coleman-de Luccia transitions in the AdS regime using holography}},
  \href{http://dx.doi.org/10.1007/JHEP09(2021)065}{\emph{JHEP} {\bf 09} (2021)
  065}, [\href{https://arxiv.org/abs/2102.11881}{{\tt 2102.11881}}].

\bibitem{Antonelli:2018qwz}
R.~Antonelli, I.~Basile and A.~Bombini, \emph{{AdS Vacuum Bubbles, Holography
  and Dual RG Flows}},
  \href{http://dx.doi.org/10.1088/1361-6382/aafef9}{\emph{Class. Quant. Grav.}
  {\bf 36} (2019) 045004}, [\href{https://arxiv.org/abs/1806.02289}{{\tt
  1806.02289}}].

\bibitem{Cuspinera:2019jwt}
L.~Cuspinera, R.~Gregory, K.~M. Marshall and I.~G. Moss, \emph{{Higgs Vacuum
  Decay in a Braneworld}},
  \href{http://dx.doi.org/10.1142/S0218271820500054}{\emph{Int. J. Mod. Phys.
  D} {\bf 29} (2020) 2050005}, [\href{https://arxiv.org/abs/1907.11046}{{\tt
  1907.11046}}].

\bibitem{Chen:2020uac}
H.~Z. Chen, R.~C. Myers, D.~Neuenfeld, I.~A. Reyes and J.~Sandor,
  \emph{{Quantum Extremal Islands Made Easy, Part I: Entanglement on the
  Brane}}, \href{http://dx.doi.org/10.1007/JHEP10(2020)166}{\emph{JHEP} {\bf
  10} (2020) 166}, [\href{https://arxiv.org/abs/2006.04851}{{\tt 2006.04851}}].

\bibitem{Witten:1981gj}
E.~Witten, \emph{{Instability of the Kaluza-Klein Vacuum}},
  \href{http://dx.doi.org/10.1016/0550-3213(82)90007-4}{\emph{Nucl. Phys. B}
  {\bf 195} (1982) 481--492}.

\bibitem{Compere:2008us}
G.~Compere and D.~Marolf, \emph{{Setting the boundary free in AdS/CFT}},
  \href{http://dx.doi.org/10.1088/0264-9381/25/19/195014}{\emph{Class. Quant.
  Grav.} {\bf 25} (2008) 195014}, [\href{https://arxiv.org/abs/0805.1902}{{\tt
  0805.1902}}].

\bibitem{Ecker:2021cvz}
C.~Ecker, W.~van~der Schee, D.~Mateos and J.~Casalderrey-Solana,
  \emph{{Holographic evolution with dynamical boundary gravity}},
  \href{http://dx.doi.org/10.1007/JHEP03(2022)137}{\emph{JHEP} {\bf 03} (2022)
  137}, [\href{https://arxiv.org/abs/2109.10355}{{\tt 2109.10355}}].

\bibitem{Banks:2000fe}
T.~Banks, \emph{{Cosmological breaking of supersymmetry?}},
  \href{http://dx.doi.org/10.1142/S0217751X01003998}{\emph{Int. J. Mod. Phys.
  A} {\bf 16} (2001) 910--921},
  [\href{https://arxiv.org/abs/hep-th/0007146}{{\tt hep-th/0007146}}].

\bibitem{Bousso:2000nf}
R.~Bousso, \emph{{Positive vacuum energy and the N bound}},
  \href{http://dx.doi.org/10.1088/1126-6708/2000/11/038}{\emph{JHEP} {\bf 11}
  (2000) 038}, [\href{https://arxiv.org/abs/hep-th/0010252}{{\tt
  hep-th/0010252}}].

\bibitem{Banks:2001yp}
T.~Banks and W.~Fischler, \emph{{M theory observables for cosmological
  space-times}},  \href{https://arxiv.org/abs/hep-th/0102077}{{\tt
  hep-th/0102077}}.

\bibitem{Witten:2001kn}
E.~Witten, \emph{{Quantum gravity in de Sitter space}},  in \emph{{Strings
  2001: International Conference}}, 6, 2001.
\newblock \href{https://arxiv.org/abs/hep-th/0106109}{{\tt hep-th/0106109}}.

\bibitem{Banks:2003cg}
T.~Banks, \emph{{Some thoughts on the quantum theory of de sitter space}},  in
  \emph{{The Davis Meeting on Cosmic Inflation}}, 5, 2003.
\newblock \href{https://arxiv.org/abs/astro-ph/0305037}{{\tt
  astro-ph/0305037}}.

\bibitem{Parikh:2004ux}
M.~K. Parikh and E.~P. Verlinde, \emph{{De sitter space with finitely many
  states: A Toy story}},  in \emph{{10th Marcel Grossmann Meeting on Recent
  Developments in Theoretical and Experimental General Relativity, Gravitation
  and Relativistic Field Theories (MG X MMIII)}}, pp.~2346--2351, 3, 2004.
\newblock \href{https://arxiv.org/abs/hep-th/0403140}{{\tt hep-th/0403140}}.
\newblock \href{http://dx.doi.org/10.1142/9789812704030_0332}{DOI}.

\bibitem{Susskind:2021dfc}
L.~Susskind, \emph{{Black Holes Hint Towards De Sitter-Matrix Theory}},
  \href{https://arxiv.org/abs/2109.01322}{{\tt 2109.01322}}.

\bibitem{Anninos:2022ujl}
D.~Anninos, D.~A. Galante and B.~M\"uhlmann, \emph{{Finite Features of Quantum
  De Sitter Space}},  \href{https://arxiv.org/abs/2206.14146}{{\tt
  2206.14146}}.

\bibitem{Morvan:2022ybp}
E.~K. Morvan, J.~P. van~der Schaar and M.~R. Visser, \emph{{On the Euclidean
  Action of de Sitter Black Holes and Constrained Instantons}},
  \href{https://arxiv.org/abs/2203.06155}{{\tt 2203.06155}}.

\bibitem{Garriga:1991ts}
J.~Garriga and A.~Vilenkin, \emph{{Perturbations on domain walls and strings: A
  Covariant theory}},
  \href{http://dx.doi.org/10.1103/PhysRevD.44.1007}{\emph{Phys. Rev. D} {\bf
  44} (1991) 1007--1014}.

\bibitem{Guven:1993ew}
J.~Guven, \emph{{Covariant perturbations of domain walls in curved
  space-time}}, \href{http://dx.doi.org/10.1103/PhysRevD.48.4604}{\emph{Phys.
  Rev. D} {\bf 48} (1993) 4604--4608},
  [\href{https://arxiv.org/abs/gr-qc/9304032}{{\tt gr-qc/9304032}}].

\bibitem{Chowdhury:2014oba}
B.~D. Chowdhury and M.~K. Parikh, \emph{{Foliation-dependence of CFTs in
  Lorentzian-AdS/CFT}},
  \href{http://dx.doi.org/10.1103/PhysRevD.93.046004}{\emph{Phys. Rev. D} {\bf
  93} (2016) 046004}, [\href{https://arxiv.org/abs/1407.4467}{{\tt
  1407.4467}}].

\bibitem{Martin:2020api}
V.~L. Martin and A.~Svesko, \emph{{Higher spin partition functions via the
  quasinormal mode method in de Sitter quantum gravity}},
  \href{http://dx.doi.org/10.21468/SciPostPhys.9.3.039}{\emph{SciPost Phys.}
  {\bf 9} (2020) 039}, [\href{https://arxiv.org/abs/2004.00128}{{\tt
  2004.00128}}].

\end{thebibliography}\endgroup

\end{document}